\documentclass[acmsmall]{acmart}

\setcopyright{rightsretained}
\acmJournal{PACMHCI}
\acmYear{2024} \acmVolume{8} \acmNumber{CSCW1} \acmArticle{11} \acmMonth{4} \acmPrice{}\acmDOI{10.1145/3637288}

\AtBeginDocument{%
  \providecommand\BibTeX{{%
    \normalfont B\kern-0.5em{\scshape i\kern-0.25em b}\kern-0.8em\TeX}}}

\usepackage{multirow}
\usepackage{tabularx}
\usepackage{xcolor}
\usepackage{url}
\usepackage{xspace}
\usepackage{caption}
\usepackage{subcaption}
\usepackage{float}
\usepackage{hyperref}
\usepackage{cleveref}
\usepackage{float}
\usepackage{soul}
\usepackage{xcolor}
\usepackage[T1]{fontenc}
\hypersetup{
    colorlinks=true,
    linkcolor=blue,
    filecolor=magenta,      
    urlcolor=cyan,
}
\usepackage{enumitem}
\usepackage{textcomp}
\setlist{leftmargin=3mm}

\usepackage[]{mdframed}

\newif\ifsubmit
\submittrue

\definecolor{Author1}{HTML}{D03416} 
\definecolor{Author2}{HTML}{13768E} 
\definecolor{Author3}{HTML}{9B4F0F} 
\definecolor{Author4}{HTML}{C99E10} 
\definecolor{Author5}{HTML}{A7D038} 
\definecolor{Issue1}{HTML}{D03416}
\definecolor{Issue2}{HTML}{13768E}
\definecolor{Issue3}{HTML}{9B4F0F} 
\definecolor{Issue4}{HTML}{008000} 

\author{Nahyun Kwon}
\email{nahyunkwon@tamu.edu}
\affiliation{%
    \institution{Texas A\&M University}
    \city{College Station}
    \state{Texas}
    \country{USA}
}

\author{Tong Sun}
\email{tsun8@gmu.edu}
\affiliation{%
    \institution{George Mason University}
    \city{Fairfax}
    \state{Virginia}
    \country{USA}
}

\author{Yuyang Gao}
\email{yuyang.gao@emory.edu}
\affiliation{%
    \institution{Emory University}
    \city{Atlanta}
    \state{Georgia}
    \country{USA}
}

\author{Liang Zhao}
\email{liang.zhao@emory.edu}
\affiliation{%
    \institution{Emory University}
    \city{Atlanta}
    \state{Georgia}
    \country{USA}
}

\author{Xu Wang}
\email{xwanghci@umich.edu}
\affiliation{%
    \institution{University of Michigan}
    \city{Ann Arbor}
    \state{Michigan}
    \country{USA}
}

\author{Jeeeun Kim}
\authornote{Co-corresponding authors}
\email{jeeeun.kim@tamu.edu}
\affiliation{%
    \institution{Texas A\&M University}
    \city{College Station}
    \state{Texas}
    \country{USA}
}

\author{Sungsoo Ray Hong}
\authornotemark[1]
\email{shong31@gmu.edu}
\affiliation{%
    \institution{George Mason University}
    \city{Fairfax}
    \state{Virginia} 
    \country{USA}
}

\ifsubmit

\newcommand{\jk}[1]{}
\newcommand{\jkIn}[1]{}
\newcommand{\xw}[1]{}
\newcommand{\xwIn}[1]{}
\newcommand{\nk}[1]{}
\newcommand{\nkIn}[1]{}
\newcommand{\ray}[1]{}
\newcommand{\rayIn}[1]{}

\newcommand{\IssueOneA}[1]{{#1}}

\else

\newcommand{\jk}[1]{\marginpar{\colorbox{Author1}{\textcolor{white}{J}} \textcolor{Author1}{#1}}}
\newcommand{\jkIn}[1]{\colorbox{Author1}{\textcolor{white}{J}} \textcolor{Author1}{#1}}
\newcommand{\xw}[1]{\marginpar{\colorbox{Author2}{\textcolor{white}{X}} \textcolor{Author2}{#1}}}
\newcommand{\xwIn}[1]{\colorbox{Author2}{\textcolor{white}{X}} \textcolor{Author2}{#1}}
\newcommand{\nk}[1]{\marginpar{\colorbox{Author3}{\textcolor{white}{N}} \textcolor{Author3}{#1}}}
\newcommand{\nkIn}[1]{\colorbox{Author3}{\textcolor{white}{N}} \textcolor{Author3}{#1}}
\newcommand{\ray}[1]{\marginpar{\colorbox{Author4}{\textcolor{white}{R}} \textcolor{Author4}{#1}}}
\newcommand{\rayIn}[1]{\colorbox{Author4}{\textcolor{white}{R}} \textcolor{Author4}{#1}}

\newcommand{\IssueOneA}[1]{\colorbox{Issue1}{\textcolor{white}{\#1A}} \textcolor{Issue1}{#1}}

\fi

\newcommand{\system}{3DPFIX\xspace}

\newcommand{\DRONE}{Easy articulation of the problem\xspace}
\newcommand{\DRTWO}{Reliable relation of a user-uploaded image with a failure type\xspace}
\newcommand{\DRTHREE}{An unobstructed flow of thoughts without being interrupted by unfamiliar technical terms\xspace}
\newcommand{\DRFOUR}{Generic solutions first, case-specific solutions on demand\xspace}

\begin{document}

\title[\system]{\system: Improving Remote Novices' 3D Printing Troubleshooting through Human-AI Collaboration}


\begin{abstract}
    The widespread consumer-grade 3D printers and learning resources online enable novices to self-train in remote settings. While troubleshooting plays an essential part of 3D printing, the process remains challenging for many remote novices even with the help of well-developed online sources, such as online troubleshooting archives and online community help. We conducted a formative study with 76 active 3D printing users to learn how remote novices leverage online resources in troubleshooting and their challenges. We found that remote novices cannot fully utilize online resources. For example, the online archives statically provide general information, making it hard to search and relate their unique cases with existing descriptions. Online communities can potentially ease their struggles by providing more targeted suggestions, but a helper who can provide custom help is rather scarce, making it hard to obtain timely assistance. We propose 3DPFIX, an interactive 3D troubleshooting system powered by the pipeline to facilitate Human-AI Collaboration, designed to improve novices' 3D printing experiences and thus help them easily accumulate their domain knowledge. We built 3DPFIX that supports automated diagnosis and solution-seeking. 3DPFIX was built upon shared dialogues about failure cases from Q\&A discourses accumulated in online communities. We leverage social annotations (i.e., comments) to build an annotated failure image dataset for AI classifiers and extract a solution pool. Our summative study revealed that using 3DPFIX helped participants spend significantly less effort in diagnosing failures and finding a more accurate solution than relying on their common practice. We also found that 3DPFIX users learn about 3D printing domain-specific knowledge. We discuss the implications of leveraging community-driven data in developing future Human-AI Collaboration designs.
\end{abstract}




\maketitle

\section{Introduction}

With the advent of technology and low-cost 3D printers~\cite{dudek2013fdm, 3d_printing_gets_bigger}, 3D printing has become increasingly available to a broader group of novice users.
Learning 3D printing means handling errors caused by several factors, such as machine settings, calibration, materials characteristics, and more. 
Handling failed printing attempts is an indispensable stepping stone to accumulating hands-on experience and being familiar with the 3D printing domain~\cite{berman2020anyone}. 
However, accommodating 3D printing and troubleshooting can be time-consuming and demanding especially for \textit{remote novices} who do not have in-person support from advanced users~\cite{berman2020anyone, hudson2016understanding, alcock2016barriers}.
Since remote novices have no in-person support, they often utilize online resources predominantly categorized into online troubleshooting archives and online communities.
Online archives are an online knowledge base that introduces comprehensive 3D printing failure types, viable potential solutions, and other online tutorials~\cite{wade2017challenges}) in a structured way (e.g., Simplify3D guide~\cite{simplify3d}).
An alternative resource is online communities, such as Thingiverse forums~\cite{thingiverse}, 3DPrinting subreddit~\cite{3dprinting_subreddit}, and Stack Overflow~\cite{stackoverflow} where group members post questions to seek help from advanced users.
While online archives present comprehensive failure types and solutions, it can be costly for novice users to build up their schematic understanding of 3D printing troubleshooting to digest information~\cite{kittur2014standing, russell1993cost}.
Meanwhile, while online communities can provide a tailored answer that can fit the individual's specific condition, finding the right solution that can work for remote novices can be uncertain. 
Also, they can easily be intolerable when the suggested solution does not work~\cite{hudson2016understanding}.
These difficulties may in turn impose a barrier for remote novices to get into the 3D printing domain~\cite{wade2017challenges, berman2020anyone}.

In this work, we improve remote novices' 3D printing troubleshooting experience through a novel human-AI collaboration design.
Our design aims at helping remote novices to naturally establish their schematic understanding~\cite{russell1993cost} about their specific printing failure types and general 3D printing knowledge through the ``learning-by-using'' approach.
To determine the requirements of our new design, we conducted a formative study (S1) with 76 active 3D printing users in the online community through an online survey. 
Based on the former work that focuses on understanding how novices use walk-up-and-print services (e.g., maker spaces, fab labs, print shops)~\cite{hudson2016understanding, berman2020anyone}, we conducted a formative study (S1) to more specifically learn how they troubleshoot using online resources, mostly online archives, and online communities.
In particular, we sought to understand how users formulate their schematic understanding of the troubleshooting task, establish their strategies, and perceive challenges and their desire for future advanced tools in leveraging online sources.
S1 found that users can access generic \& common solutions listed in online archives, but novices had more difficulties in understanding how they can lexically explain their problem when searching.
When matching their case with representative images provided for each printing failure type, their limited understanding of 3D printing and unfamiliar technical terms further constrained them.
Meanwhile, we found novice users tend to rely on online communities to get custom solutions as they can inquire easily with plain languages (e.g., photos of a failed print and print settings). 
But there exists inevitable latency to get the answer mainly due to the unbalanced number of novices seeking help and experts who can help.
Rather than doing their ``homework'' by reading the cumulative knowledge in the online community, we found users tend to upload the issues that have been explained in the past.



 

Based on the insights found in S1, we designed \system, a system that provides an AI-driven ``curated'' troubleshooting workflow in detecting the failure type and finding a solution.
In building our design, \system transfers \textit{\textbf{social annotation}}~\cite{kittur2014standing}---the knowledge accumulated in online communities contributed by their community members---into intelligent and interactive troubleshooting guidance powered by AIs to significantly reduce novices' effort for accomplishing troubleshooting.
First, to leverage social annotation in designing \system, we extracted information accumulated in the 3D printing community relevant for automating the Q\&A process, such as user-posted photos showing the failed print and textual discussion threads to diagnose/resolve the failure and implement failure type classification models.
This design decision was made to help remote novices articulate their cases more easily by simply uploading their photos, instead of relying on domain-specific terminologies to describe their cases for search. Especially in remote settings, using visual cues can lead users to self-investigate by helping them gain a deeper understanding of the issue (e.g., a certain setting can cause certain visual cues on print). This would be more useful in remote settings than just using current text-based search schemes.
Second, \system helps users better relate their image with possible failure types suggested by AIs, \system leverages the XAI technique, normally known as a local explanation along with a structured explanation of visual features of the failure type collected from the online community.
Third, upon the selection of the failure type, \system suggests the feasible solutions, starting from the most generic solution and then advanced solutions based on a user's look-up.

To understand how the new design can improve remote novices' experience in troubleshooting, we conducted a summative study (S2) with 13 novice users in experimental settings
and 6 users in more natural settings.
By deploying \system for a short period for qualitative feedback, we received comments from 6 active 3D Printing users.
S2 found several positive signals of \system in helping remote novices.
For example, S2 found using \system helped participants find a solution with significantly reduced effort and cognitive load. Their perceived efficiency in diagnosing the failure type was significantly lower than their common practice.
The solutions the participants found were evaluated by 3D printing experts in a blind setting.
As a result, expert reviews found that the solutions they found using \system were rated significantly higher than the baseline scores.
Finally, \system significantly increased learnability-related performance than baseline.

This work offers the following contributions:
\begin{itemize}
    \item \textbf{S1: Understanding Remote Novices' 3D Printing Troubleshooting Practice and Challenges}:
    We seek to more deeply understand how remote novices utilize online resources in their troubleshooting practice and what challenges they encounter in general.
    Based on our findings, we derive design requirements that can improve remote novices' troubleshooting experience.
    \item\textbf{\system}:
    We build \system, a novel human-AI collaboration design devised based on S1 findings.
    Using the collective social annotations extracted from online archives and communities, \system realizes a series of image classifiers for supporting remote novices' detection of their failure type and the mapped solutions.
    \item \textbf{S2: Effect of \system}:
    We compare remote novices' use of \system to their current practice to understand how \system can positively impact novice users' task performance and 3D printing learnability in both behavioral and perceptual manners.
    \item \textbf{Implications for Design}:
    We reflect on the implications of considering social annotation in developing AI-driven troubleshooting systems to improve users' 3D printing troubleshooting experience and beyond.   
\end{itemize}

\newpage
\section{Related Work}

We briefly cover empirical studies that deepen our understanding of how novices learn 3D printing techniques and knowledge in situ.
Several previous studies have focused on 3D printing novices who have in-person support, which can largely help them to be familiar with the 3D printing domain~\cite{hudson2016understanding}.
As the makerspace concept became widespread through public print centers, fab labs, and creating hubs, several studies found that in-person help enables casual makers who lack prior knowledge in 3D printing to have advanced experiences in operating and maintaining machines~\cite{berman2020anyone}.
In general, however, getting support from experts is not easy for casual makers as 3D experts are scanty~\cite {hudson2016understanding}.
Meanwhile, the blooming internet culture has enabled people in remote environments to get fast access to social contracts through online communities~\cite{rayna2015co}.
Online communities play a key role in letting 3D printing newcomers gradually adopt new concepts through social interaction with the more experienced practitioners~\cite{gray2005informal}.
However, printing 3D objects at home without any in-person support from experts can be finicky.
One example can be adjusting print parameters on their machine from shared 3D objects online~\cite{oehlberg2015patterns}.
Since the novices lack prior knowledge to create 3D objects due to difficulty in learning 3D modeling software, they often download user-created 3D files from public repositories such as Thingiverse \cite{thingiverse, hudson2016understanding, alcock2016barriers}. 
Since not every 3D modeler shares the detailed print settings including the machine or material used \cite{oehlberg2015patterns} that are often critical for successful 3D printing, it costs more time for novices to find the optimal settings.
A study also pointed out that it is common for the modelers not to update changes, and even if they do, updating modification is often done in the comments that could be easily overlooked \cite{tseng2014product}.

Next, we review work in sensemaking and social annotation to characterize the nature of troubleshooting as an exploratory cognitive task. 
Sensemaking is the process of establishing one's internal representation of the target problem space through interacting with information resources~\cite{russell1993cost, pirolli2005sensemaking}. 
Several theories have been proposed to explain a user's behavior in sensemaking, including cost structure model~\cite{russell1993cost}, information foraging~\cite{pirolli1995information}, and more~\cite{thomas2005illuminating}.
At an early stage, sensemaking focused on an ecosystem between a single user and a single information system.
As the literature in CSCW and social computing grows, sensemaking has influenced on developing social information foraging theory~\cite{pirolli2009elementary}.
Social information foraging has largely influenced in designing CSCW applications, including collaborative information seeking~\cite{hong2018collaborative, hong2019design}, collaborative data analytics~\cite{tang2006collaborative}, collaborative search~\cite{morris2010wesearch}, social network question and answering~\cite{morris2010people}, and beyond~\cite{thomas2005illuminating, chung2019efficient}.
3D printing troubleshooting is achieved largely by active interaction between a novice and an advanced users in a remote setting using the information repository built based on social annotation~\cite{kittur2014standing}.
On top of the established social annotation, novices achieve their sensemaking goals by defining search keywords, reading articles, and refining keywords to follow up to gradually develop one's internal representation.
Through their exploration, they may be acquainted with domain-specific knowledge and language, types of viable 3D printer models and their pros and cons, and more importantly, types of printing failures and plausible reasons about why they happen, how to fix them, and many more~\cite{alcock2016barriers}.

While social sensemaking and annotation have inspired numerous applications in CSCW, there have been relatively a scarce of approaches that leverage AI in facilitating novices' sensemaking process, e.g., ~\cite{heer2019agency, chuan2022flatmagic, hong2020human, santos2019visus, hong2019disseminating} in troubleshooting.
Rather, many focused on improving 3D printing computational pipelines, such as real-time failure detection systems with video cameras.
A quality monitoring pipeline can compare the location or shape of the 3D object in the middle of printing using machine learning techniques~\cite{delli2018automated, baumann2016vision, paraskevoudis2020real}.
In practitioners' fields, AI Build ~\cite{aibuild_article} caters an autonomous, large-scale 3D printing with real-time failure detection and correction in robot-arm-based printing, powered by computer vision technology.
However, these approaches are scaffolded by additional hardware support that enables real-time monitoring and capturing failures, which might not be an optimal solution for end-users~\cite{delli2018automated, baumann2016vision, paraskevoudis2020real, aibuild_article}.
This monitoring technique with a camera also tends to be bound to the lighting condition, the specific printer they used for experiments~\cite{delli2018automated}, and calibration for an accurate camera positioning~\cite{baumann2016vision}.
In general, previous AI-driven techniques aim at detecting machine-originated defects rather than supporting human operators' understanding.

Through our review, we identify the two gaps we can investigate further to facilitate remote novices' learning of 3D printing through their troubleshooting experience.
First, while several studies have provided useful insights regarding how 3D printing users motivate their learning through in-person setting~\cite{hudson2016understanding, erden2012knowledge,rayna2015co}, we are motivated to further understand how novices use online resources in coping with troubleshooting in remote settings. 
Therefore, we further investigate practices and challenges of the remote novices in utilizing popular online troubleshooting guides and soliciting support from advanced users to identify new design opportunities.
Second, while social information foraging and social annotation have provided a useful framework that designers can use to transfer their insights into new applications, developing an interactive human-AI collaboration design for 3D printing remote novices has not been seriously considered in the past literature. 
To better support our target users, we adopt a user-centered design to develop a viable system that motivates lowering barriers for 3D printing troubleshooting.

\section{S1: Formative Study}\label{section:s1}

Taking one step further from the previous work that focused on the novices who have in-person support~\cite{hudson2016understanding, berman2020anyone}, Study 1 (S1) aims at two main objectives:
\begin{itemize}
    \item Seeking to understand \textit{remote} novices' practices, challenges, and future desire in handling 3D printing troubleshooting.
    \item Establishing design requirements for support tools that can potentially improve remote novices' troubleshooting processes.
\end{itemize}

As a starter, we crawled 26,894 posts from the FixMyPrint subreddit~\cite{fixmyprint} to understand how 3D printer users communicate for troubleshooting in online communities. Next, we conducted an online survey to understand novices' strategies, challenges, and desires in utilizing existing online resources for troubleshooting compared to user groups with advanced knowledge.

\subsection{Preliminary Observation} \label{section:preliminary}

\begin{figure}[!t]
    \centering
    \vspace{-0.5cm}
    \includegraphics[width=1.0\linewidth]{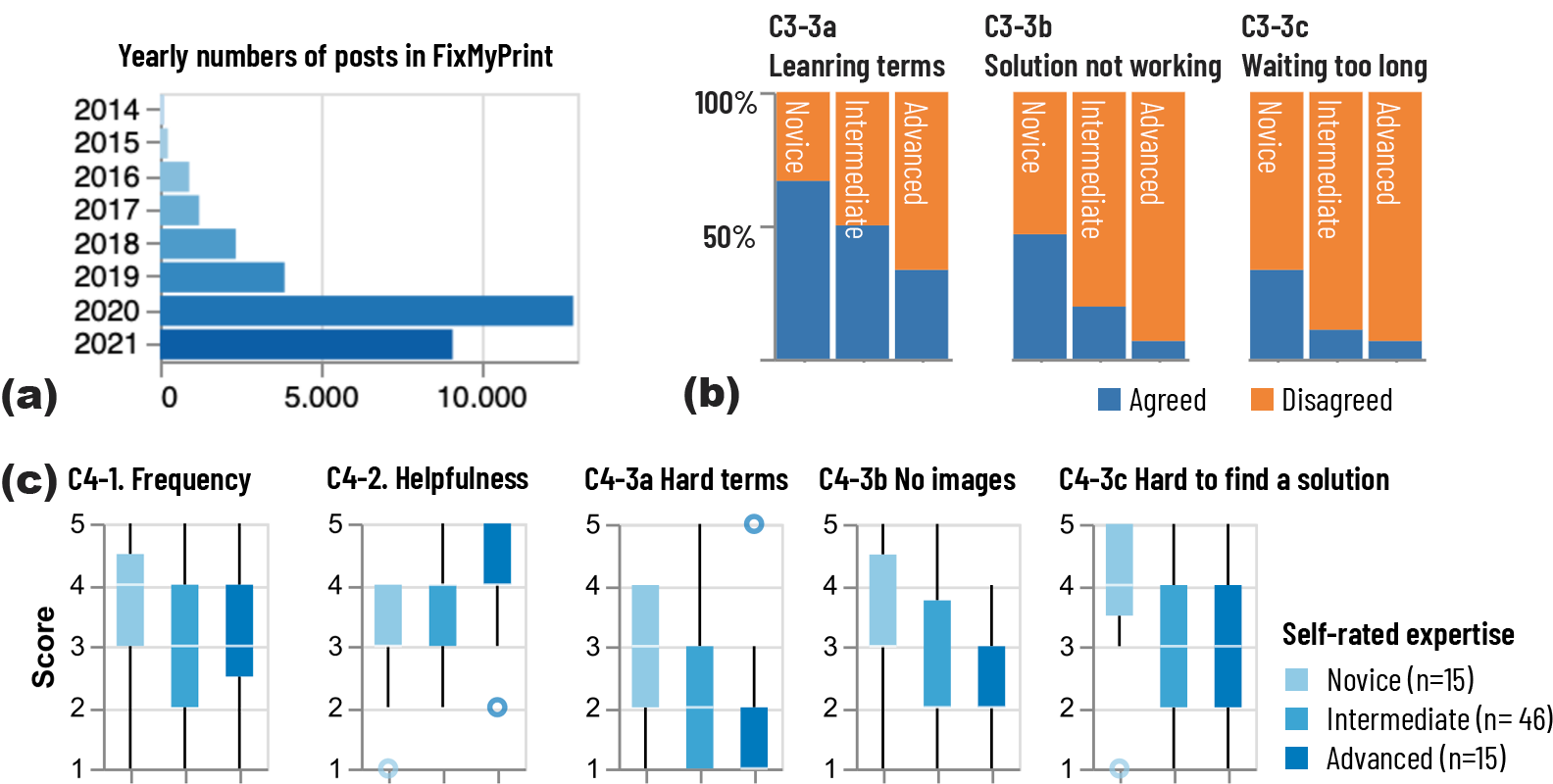}
    \caption{(a) Yearly count of Q\&A posts in FixMyPrint (2021 counts posts between Jan. and June), (b) Survey results of C3-3, users' changing perception regarding online communities depending on their self-rated expertise, (c) Survey results of C4-3, users' changing perception regarding online archives depending on their self-rated expertise}
    \Description{}
    \label{Fig:S1}
    \vspace{-0.5cm}
\end{figure}

In this phase, we observed behaviors of members in FixMyPrint subreddit~\cite{fixmyprint}, one of the popular online 3D printing troubleshooting communities with over 100k members, to understand how they share 3D printing issues and interact with others to solve them using a board discussion. 
We also looked into FixMyPrint's policy \& announcements to see how they foster effective troubleshooting-related communication.
Our observation revealed the growing need for targeted solutions for novices, continuously growing posts seeking answers to fix their issues (see Figure~\ref{Fig:S1}, (a)). 
Such needs seemed to be aligned with the active participation of motivated 3D printing enthusiasts and experts' altruistic motivations to benefit 3D printing novices as a community.

Unfortunately, a nontrivial amount of posts are eventually left unanswered.
One viable reason is that questions are repetitively overflowing the boards. 
Many of them share commonalities thus existing solutions can be easily found in prior posts by searching and finding similar cases, which is a practice that is not well-used by novices.
The announcement on the main page recommends reading online web archives first before posting questions, especially Simplify3D guide~\cite{simplify3d} first, along with other options such as Rigid.ink~\cite{rigid_ink}  Matterhackers~\cite{matterhackers}, and Reprap~\cite{reprap} if desired.
AutoModerator in this community immediately sends an automatic comment for every new post saying \textit{``Most common print quality issues can be found in the Simplify3D print quality guide''}.
Many online communities' internal recommendation system pulls popular posts onto the main page, further marginalizing non-busy posts and depreciating ones with only a little to no discussion.
In our informal conversation with community moderators, they seemed to be aware of this issue, encouraging users to re-post when initial attempts were left unresolved due to a lack of attention.

As Figure~\ref{Fig:S1} (a) shows, demand for seeking a targeted answer is soaring while experts' availability to read and comment remains flat, resulting in an imbalance. 
Among $\approx27k$ posts that we collected from the FixMyPrint subreddit, 
15\% of the posts ($\approx6k$) have never been answered and nearly 50\% have 3 comments or fewer.  
Given the depth of discussion needed to reach an eventual solution, our data acknowledge substantial posts ending up being unsuccessful attempts.
For posts with at least one comment, the average time to get the first comment was about 789 minutes, which indicates the wait time for getting attention is nearly 13 hours on average, which even does not guarantee resolving the issue.
We assume such an imbalance in supply and demand can hinder novices from finding valid solutions and attaining 3D printing domain knowledge timely.
Through our observation, we were motivated to deeply understand the challenges of novices, as well as to identify further how the current online archives impose more challenges.

\subsection{Online Survey}

\subsubsection{Methodology}
As a way to understand the challenges of 3D printing users in troubleshooting failures more deeply, we designed a survey and reached out to active 3D printing practitioners in online communities.
To recruit 3D printing users online with different levels of expertise, we posted invitation flyers to popular Reddit 3D printing communities including FixMyPrint~\cite{fixmyprint}, Ender3~\cite{ender3_subreddit}, 3DPrinting~\cite{3dprinting_subreddit}, and Prusa3D~\cite{prusa_subreddit} that had at least 40k members. 
Then we reached out to 842 members of the chosen subreddits who recently posted inquiries through direct messages.
We also sent an invitation flyer to popular 3D printing-related discord channels with more than 2k members, titled Print Everything~\cite{print_everything_Discord}, Creality 3D Printers~\cite{creality_discord}, and Prusa3D~\cite{prusa3d_discord}.
Our invitation resulted in 76 responses being returned.

In the survey, we first asked participants to self-evaluate their expertise in 3 levels: novice ($G_{nov}$), intermediate ($G_{int}$), and advanced ($G_{adv}$). 
We used self-assessed expertise instead of more quantitative measures such as years of experience. Unlike other domains such as education that can fairly reflect people's expertise using quantitative metrics, 3D printing is highly arbitrary in defining the expertise. 
For example, many hours of taking workshops and being tightly committed and engaged in makerspaces may intensively increase their expertise in a short time.
Among 76 respondents, 15 self-assessed themselves as novice while 46 assessed as intermediate and 15 assessed as advanced, respectively. 
Then we asked the 9 questions of the 5 categories below:

\begin{itemize}[leftmargin=.3in]
    \item[\textbf{C1.}] \textbf{Printing environment}: What type of filaments and printers do you use?
    \item[\textbf{C2.}] \textbf{Common strategies for troubleshooting}: Please indicate your ``strategies'' to use varying online resources when handling troubleshooting.
    \item[\textbf{C3.}] \textbf{About using online communities}
        \begin{itemize}[leftmargin=.3in]
            \item[C3-1.] When using online communities, what do you usually do to resolve issues?
            \item[C3-2.] Which aspects of online communities help you the most?
            \item[C3-3.] Please indicate if you agree with the statements as follows: When using online communities, (a) I learned technical terms and several useful tips through discussion; (b) Solutions suggested did not really solve my issue; (c) I needed to wait too long to get others' responses.
        \end{itemize}
    \item[\textbf{C4.}] \textbf{About using online archives}
        \begin{itemize}[leftmargin=.3in]
            \item[C4-1.] How often do you read online archives?
            \item[C4-2.] To what extent is reading online archives helpful in resolving your issues?
            \item[C4-3.] Please indicate how much you agree with the following aspects in applying information learned from online archives: When using online archives, (a) It is hard to understand technical terms; (b) None of the example images in the archives are similar to my case; (c) It is hard to find a specific solution that applies to my own printing issue
        \end{itemize}        
    \item[\textbf{C5.}] \textbf{Further remark}: Open-ended questions that they can freely describe their experience regarding troubleshooting using online resources.
\end{itemize}
For questions in C1-3, we provided multiple choice options that respondents can choose, including `other' to describe further answers that do not fall under any options. 

\subsubsection{Results}

About the common troubleshooting strategies (C2), reading well-known online archives and the most popular suggestion (\textit{N} = 61, 79\% of participants), followed by searching previous online community posts (64\%), and posting problems on online communities' (51\%). 
Relatively fewer respondents relied on video sources such as YouTube (11\%) or search Google (9\%).

\textbf{Online Communities: difficulty in searching and obtaining timely help.}
We designed C3 to confirm the usage patterns we found in the preliminary online community observation and to hear more about barriers and desires.
It seemed that users post their own questions since searching the key articles often fails, regardless of their level of expertise.
P32 (intermediate) mentioned: \textit{``First of all, [I] look for previous posts with similar issues, then ask those that answered those posts, and if nothing worked, posting my own with details on settings [...] and where it failed with a close-up on the failure''}.
The most common troubleshooting strategy (C3-1) was searching for posts that handled a similar problem from the previous discussion board (69.2\%) followed by uploading posts with the photos of failed prints (62.8\%).
Many understood how useful well-articulated discussion records with details are to obtain more accurate suggestions, as well as the fact that they assist by building assets and a healthy community attracting more members in the longer term.
Many respondents indicated that their initial posts include sufficient details: \textit{``I have found giving as much detail along with posting images is a good way to not only get help myself but provide people in the future a resource as well (P14, intermediate)''}.

Nonetheless, several novice participants found it hard to search for the right discussion record that matches their case. For instance, P19 (novice) pointed to the problem of Reddit's search function: \textit{``Reddit's decentralized model limits search functionality because of lack of tags or use of key terms''}.
P17 (novice) mentioned that \textit{``Searching doesn't always help, but I try it anyway just to do due diligence''}.
Not knowing the terminology to articulate what is happening, novices' success rate to hit the right post discussing similar cases through text-based query has no choice but to be limited.

Several respondents found getting comments from others in online communities hard (C3-3), while the majority of respondents find immediate support is the key.
61.5\% of respondents noticed that the benefit of online communities significantly reduces when they cannot have ``synchronous feedback'' or ``instant support''.
\textit{``The only real issue is that it can take quite a while to get just one response, and sometimes posts end up not getting answered at all''} (P26, intermediate).


As found (Section ~\ref{section:preliminary}), the core value of online communities seemed custom feedback, obtained from experienced users without domain-specific language. 
61.5\% of respondents loved using online communities for troubleshooting owing to the possibility of getting their issues diagnosed using photos of a failed print.
Meanwhile, people with lower expertise felt they learned more about 3D printing language than people with higher expertise did (see Figure~\ref{Fig:S1}, (b) C3-3a).
On the other hand, the risk of using online communities is the supply-and-demand gap in Q\&A, which inevitably disables synchronous feedback and lowers the chance of getting a timely solution.
Interestingly, we found novice users perceive the waiting time as longer than more advanced groups (see Figure~\ref{Fig:S1} (b) C3-3c).
For instance, P10 (intermediate) drew attention to the high demand of novice users requesting help in comparison with the relatively low number of experienced users giving useful suggestions, stating \textit{``There are too many users requiring help compared to those who can actually help. That makes it harder for me to get a response''}.
Even though the perceived waiting time of the advanced user groups is relatively shorter than the novice group, they are aware of the issue; P13 (advanced) criticizing Reddit's availability noted; \textit{``Reddit is not an appropriate forum for troubleshooting. Troubleshooting should really be done with live [and more interactive] feedback. Posting a bunch of pictures and waiting doesn't lead to good results''}.

In sum, we found discourses of online communities and custom suggestions from experienced users benefit remote novices.
They learn domain-specific languages and experience useful tips to fortify their 3D printing knowledge by obtaining suggestions from others, trying recommended solutions, and asking back for further help if not resolved. However, we also found experienced users in online communities scarce, making it hard for novices to elicit suggestions.

\textbf{Online archives: unfamiliar technical terms, difficulty in finding applicable problem \& solution.}
C4's main goal is to understand how and why users' perceptions about using online archives vary, and whether there exists differences depending on their expertise.
We conducted Spearman's rank correlation for every C4 question with the null hypothesis that respondents' Likert scale rates will not vary between groups with different expertise levels ($H_0$).
Regarding the frequency of using online archives (C4-1), Spearman's rank correlation test found that the frequency didn't vary significantly depending on their level of expertise ($G_{nov}$: \textit{M} = 3.53, \textit{SD} = 1.20, $G_{int}$: \textit{M} = 3.27, \textit{SD} = 1.18, and $G_{adv}$: \textit{M} = 3.27, \textit{SD} = 1.18, r(74) = 0.036, p = 0.754, see Figure~\ref{Fig:S1} (c)).
On the other hand, we found that people with lower expertise tend to perceive online archives significantly less useful than the group(s) with more experience ($G_{nov}$: \textit{M} = 3.07, \textit{SD} = 0.85, $G_{int}$: \textit{M} = 3.93, \textit{SD} = 0.77, and $G_{adv}$: \textit{M} = 4.13, \textit{SD} = 0.81, r(74) = 0.449, p = 0.000, see Figure~\ref{Fig:S1} (c)). We especially noticed the gap between $G_{nov}$ and $G_{int}$ to be way larger than the gap between $G_{int}$ and $G_{adv}$. 
These results indicate that the level of expertise does not affect the frequency of use, but novices may be unlikely to obtain useful information, compared to advanced users.

Regarding the analyses to understand possible factors that make the online archives hard to use, Spearman's rank correlation rejected every null hypothesis in this category. 
First, technical terms can be a bigger barrier to using online archives for novices than the experienced ($G_{nov}$: \textit{M} =  3.40, \textit{SD} = 1.20, $G_{int}$: \textit{M} = 2.65, \textit{SD} = 1.07, and $G_{adv}$: \textit{M} = 2.47, \textit{SD} = 0.96, r(74) = -0.250, p = 0.030, see Figure~\ref{Fig:S1} (c), C4-3a). 
Next, Spearman's rank correlation test found that relating example images used in the online archive to a user's specific problem becomes harder if their expertise levels are lower ($G_{nov}$: \textit{M} =  2.87, \textit{SD} = 1.09, $G_{int}$: \textit{M} = 2.11, \textit{SD} = 1.07, and $G_{adv}$: \textit{M} = 1.80, \textit{SD} = 1.11, r(74) =  -0.313, p = 0.006, see Figure~\ref{Fig:S1}, C4-3b). 
Finally, the lower self-rated expertise levels are, the more difficulties in applying a general level of solution description to their specific case ($G_{nov}$: \textit{M} =  4.00, \textit{SD} = 1.10, $G_{int}$: \textit{M} = 2.93, \textit{SD} = 0.92, and $G_{adv}$: \textit{M} = 3.07, \textit{SD} = 1.18, r(74) =  -0.269, p = 0.019, see Figure~\ref{Fig:S1} (c), C4-3c).
These findings could explain why online archives are perceived as less useful by users with lower-level expertise particularly. 
Novices may need to invest more time and effort to understand domain-specific language to fully digest the articles. 
Limited knowledge about terminologies may also negatively affect the way that novices search posts due to the text-based service.
While images are more direct and intuitive to deliver the context, representative images displayed may be unfamiliar, making it also hard to locate similarities given every single sign and symptom could be unique in different users. 
Even if they were able to find an article relevant, novices may also face barriers in selecting one among several suggested solutions.

\subsection{Design Requirements for Remote Novices}\label{section:dr}

Remote novices' desire seems straightforward--(1) being capable of easily identifying their problems (2) using an easy inquiry, and (3) obtaining the targeted solution to resolve the issue directly, without circumvention or prediction.
``Remote experts'' are a big help since they can be immediate in noticing issues that are shown with minimal textual/visual information, and give straightforward answers to seemingly-unique printing issues than examining existing posts.
However, the overwhelmingly increased volume of posts makes the experts an uncertain asset.

To mitigate the gap we identified in S1, we propose a novel human-AI collaboration system leveraging social annotation obtained from existing online communities that have accumulated remote experts' knowledge over time, which can advance remote novices' 3D printing troubleshooting experience. 
Using social annotation (comments), we seek to expand the role of AI as a curator who can balance the load for online experts. 
The role of these remote experts includes guiding novices to identify the type of problem, match feasible solutions in plain language, enabling access to targeted solutions similar to the human experts' custom solutions.
Specific design requirements we learned from\ S1 are shown as follows.

\textbf{DR1. \DRONE}:
Troubleshooting often starts by investigating noteworthy visual features from a failed print to understand the printing failure type.
Visually investigating the problem is an intuitive and straightforward way to diagnose a failure type.
To facilitate easy diagnosis, our first DR is to use photos as input.
By enabling an image-based search, remote novices can avoid describing unfamiliar domain terms.
Considering text-based communication appeared in both online archives and communities, we hypothesize that an automatic diagnosis powered by vision-based models can better assist remote novices.

\textbf{DR2. \DRTWO}:
Diagnosing 3D printing failures type by referring to a single representative image for each type could not be complete, because
failures visual traits can be different from one case to another.
To help remote novices better identify the most viable failures shown in the image, the system should help users get the visual reasoning of the failure types (e.g., \cite{choi2019aila, gao2022res, gao2021gnes}). Through the system, users should be able to understand which visual features belong to a certain failure type. 

\textbf{DR3. \DRTHREE}:
The use of synonyms without explicit clarification can confuse users 
(e.g., stringing - wisping, vibration - ghosting/rippling/ringing), 
the new design should provide an explanation of newer technical terms quickly.
Being on-demand is also essential considering the different levels of knowledge that user groups might possess.

\textbf{DR4. \DRFOUR}:
Since several different reasons underlie the seen features of printing failures (e.g., too low printing temperature and/or high printing speed can cause under-extrusion), there could be a diverse set of how-to suggestions. 
Also, searching for previous discussions is a common behavior to find useful dialogues but one-size fits all does not apply to this case. 
A built-in search functions are limited to retrieving the applicable posts, and a lack of novices' expertise may result in the wrong choice of suggestions.
Improving the capability to find the applicable post among existing ones is thus essential,
and to help users select the highly applicable solution which is worth trying first among many alternatives will eventually help reduce the time for trial and error.
To address challenges in using two online resources, we apply the visual information-seeking mantra, ``overview first, zoom and filter, and details on demand''~\cite{shneiderman2003eyes} in providing feasible solutions.
Online archives list generic solutions along with the description of the failure itself and common reasons that cause such problems.
Also as the online communities tend to encourage novices to append context (e.g., printer and filament types and settings) for accurate diagnosis and case-specific solutions, we are to provide a detailed description of the diagnosed failure and the generic solutions first (overview). If desired, users can proceed to open more case-specific solutions that do not fall into the generic cases (details on demand).
Also, if available for individual solutions, the system should provide conditions or clues that further narrow down the search for options (zoom and filter). 

\section{\system: AI-driven Design for Automated 3D Printing Troubleshooting}

Online communities have accumulated ample real-world image data on 3D printing failures and discussions as a form of social annotation.
While such resources provide a significant amount of textual records about the issues showing in the images, i.e., \textit{social annotations}, remote novices' process to get the solution is not without uncertainty as per our S1 findings.
For instance, many of their postings may not receive any attention at all, they may have to wait until other experienced users respond, or they may overlook the existing posts and make the same question.
Using static online archives is also challenging due to unfamiliar technical terms, difficulties in relating their case with the description, and selecting suggestions to try. 
To provide a better troubleshooting experience to remote novices, \system provides interactive features built based on the DRs we derived in S1.
In particular, it provides an automatic diagnosis of the 3D printing failure type suggested by AI and helps users find corresponding solutions based on social annotations that we extracted from 3D printing online archives and communities.




\subsection{\system Design Overview \& Implementation} \label{system_overview}

Following the four design requirements defined in S1, \system is equipped with four main features: (1) a diagnosis of failure types from the user-uploaded photo (\textbf{DR1}), (2) visual reasoning through the grayscale saliency attention map to provide a local explanation of the CNN model's decision and representative sample images (\textbf{DR2}), (3) a detailed explanation of technical term with a hover-over window on demand (\textbf{DR3}), and (4) generic \& case-specific solutions with filter conditions to narrow down the search (\textbf{DR4}).

\begin{figure*}[!b]
    \centering
    \includegraphics[width=\linewidth]{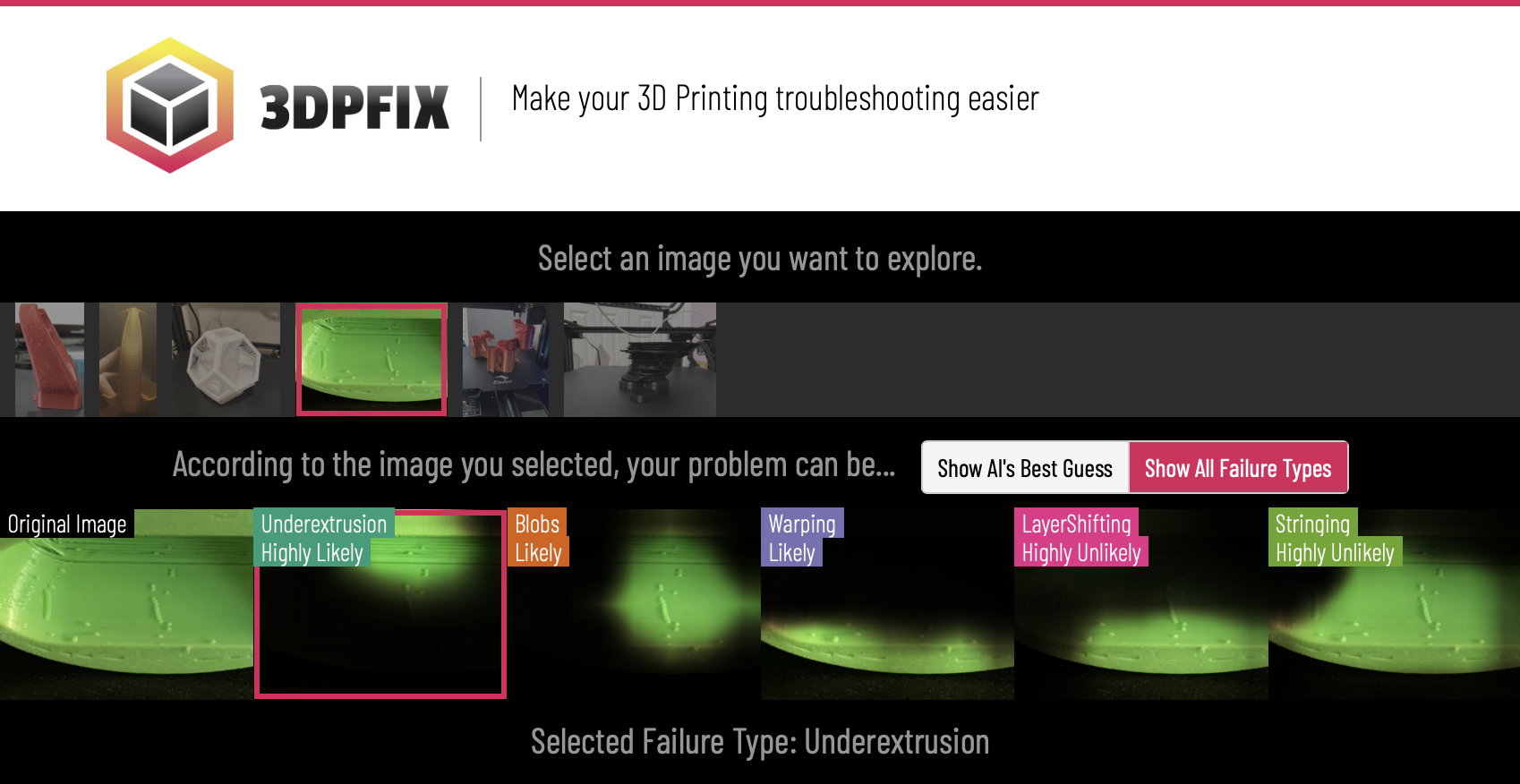}
    \vspace{-0.5cm}
    \caption{This is the top section of the interface where users can select the 3D printing images they uploaded for diagnosis (upper subsection) and the corresponding failure type predictions with saliency maps generated by our models (lower subsection). Images are clickable tabs where red-colored borders indicate active selections. The system also displays the likelihood of each failure type prediction (Highly Likely: 75\% - 100\%, Likely: 50\% - 75\%, Unlikely: 25\% - 50\%, and Highly Unlikely: 0\% - 25\%). Users can click a specific failure prediction to explore further the solutions on the bottom section of the interface shown in Figure~\ref{figure:system_description}, \ref{figure:system_solutions}, \ref{figure:system_hoverover}. Users can also use the toggle switch buttons to ``Show All Failure Types'' or ``Show AI's Best Guess'' which filters out the predictions below the `Highly Likely' level.}
    \Description{}
    \label{figure:system_diagnosis}
\end{figure*}


\textbf{\#1. Image-based diagnosis for the easy articulation of the problem (DR1):}
Visual characteristics of failed prints play a key role in recognizing 3D printing failure types; automated diagnosis through user-uploaded photos is the most intuitive and simple way for novices.
Users can start by uploading photos of their failed prints using our web-based interface.
\system examines if it contains any type of 3D printing failure by computing its probability of presence using all CNN classification models step by step.
\system shows results within a few seconds and lets users directly recognize in synchronous settings.
\system does not display prediction probabilities directly to the users as presenting 
numeric values can confuse users possibly due to an objective interpretation~\cite{xie2020chexplain}. 
Instead, \system presents abstracted labels of prediction results in four levels: Highly Likely (75\% - 100\%), Likely (50\% - 75\%), Unlikely (25\% - 50\%), and Highly Unlikely (0\% - 25\%). 
As in Figure~\ref{figure:system_diagnosis}, the `Show AI's Best Guess' is activated which shows the `Highly Likely' failure type visible only by default. 
Upon preference, users can also enable `Seeing all failure types' to check all the failure types that \system supports.

\begin{figure*}[!b]
    \centering
    \includegraphics[width=\linewidth]{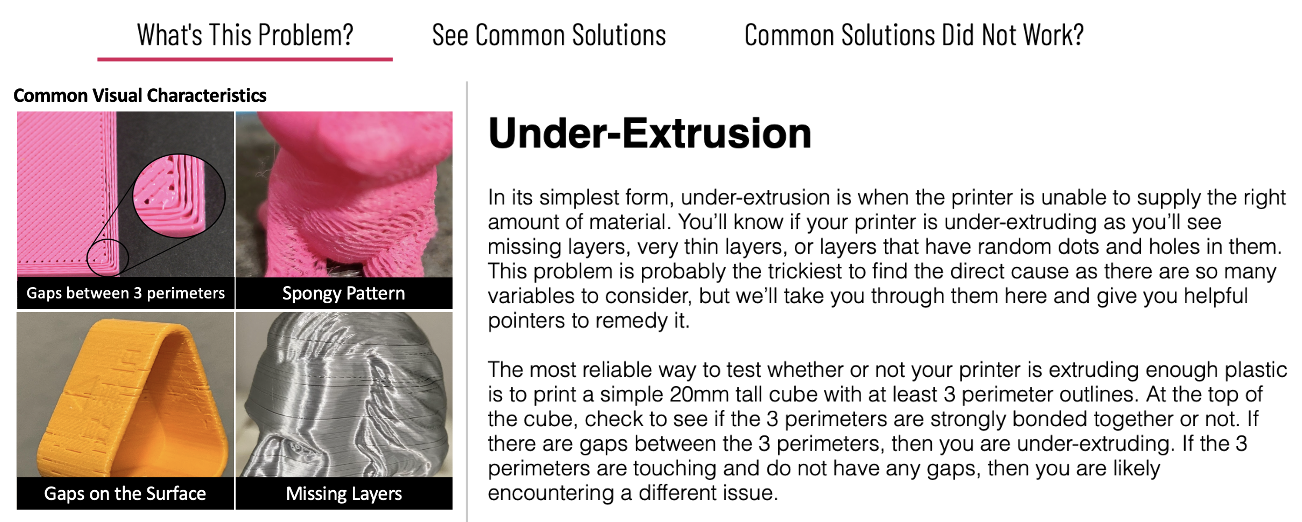}
    \vspace{-0.5cm}
    \caption{The bottom section of the interface has 3 tabs to investigate solutions for the selected failure type by the module in Figure~\ref{figure:system_diagnosis}. The first tab, `What's This Problem?' shows example photos about the common visual characteristics of a failure type, as well as an easy-understandable description on the right.}
    \Description{}
    \label{figure:system_description}
    \vspace{-0.5cm}
\end{figure*}

\begin{figure}[!t]
    \centering
    \includegraphics[width=0.6\linewidth]{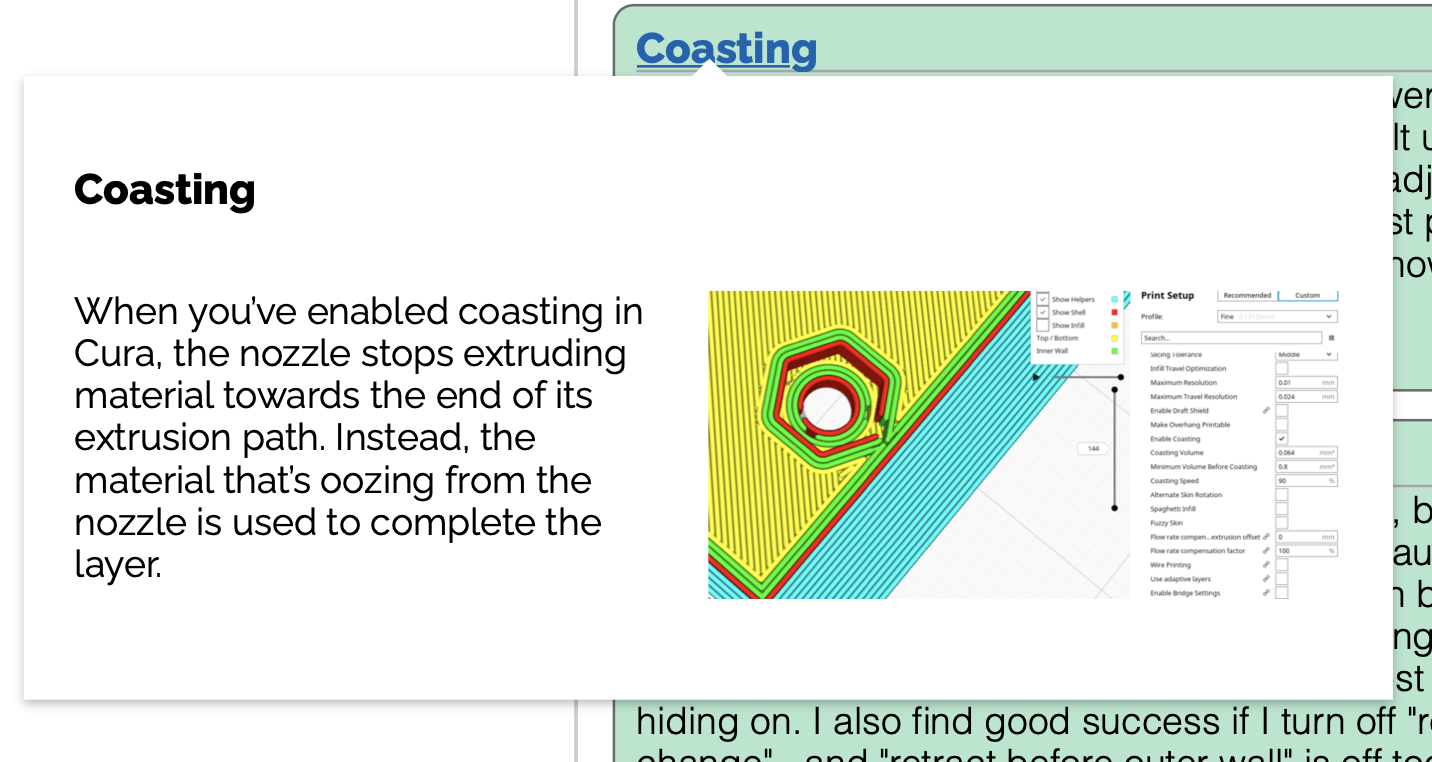}
    \vspace{-0.5cm}
    \caption{Some technical terms in the solutions have detailed explanations and example photos when users hover over the terms (blue-colored with underlines). These underlined terms are also clickable on external websites with more comprehensive descriptions.}
    \Description{}
    \label{figure:system_hoverover}
    \vspace{-0.5cm}
\end{figure}

\textbf{\#2. Visual reasoning of diagnosis for reliable relation between user-uploaded photos and diagnosed failure types (DR2):}
In addition to the diagnosis, \system provides visual reasoning on the diagnosed results, which can support users to not only better appreciate classification models' decisions, but also learn various visual features to recognize different types of failure.
As a lower subsection of Figure~\ref{figure:system_diagnosis} shows, our tool also provides the grayscale saliency map generated by Grad-CAM \cite{selvaraju2017grad} for each type diagnosed.
Saliency maps present the region used for prediction, highlighting the point of interest. 
This accentuation helps users focus on the problematic parts in the photo, as well as contribute to the visual explanation of CNN models where it is used to make the decision~\cite{preece2018asking}.
Upon clicking each failure type, \system provides the detailed descriptions under the \textit{`What's this problem?'} tab.
This tab contains several representative images with different visual features, extracted from our image dataset (See Figure~\ref{figure:visual_feature}), which will provide another visual reasoning for users to understand why such failure types were diagnosed.
These images assist users to discern common visual characteristics, facilitating learning about print failures with their appearances.
In addition, users can also read textual explanations for each failure type, such as what it is called in the 3D printing community (jargon), what can cause it, and which visual characteristics commonly appear.
Here users are able to learn about the type of failure, terminologies, visual characteristics, and the clue (what causes them), users move to the next step to learn about solutions.



\textbf{\#3. Hover-over functions for a detailed description of technical terms (DR3):}
In the 3D printing domain, there are many technical terms that refer to specific settings or various techniques to improve printing quality. 
Remote novices often face barriers due to unfamiliar technical terms during their troubleshooting processes, which makes them switch to searching for external sources.
To minimize such intervention, as in Figure~\ref{figure:system_hoverover}, \system uses hover-over boxes, as many encyclopedic web services such as Wikipedia adopt, to further aid newcomers learn technical terms used in the text when describing solutions.
According to their knowledge levels, users can easily hover over the terms on-demand and read a description along with the image.

\begin{figure*}[!b]
    \centering
    \includegraphics[width=\linewidth]{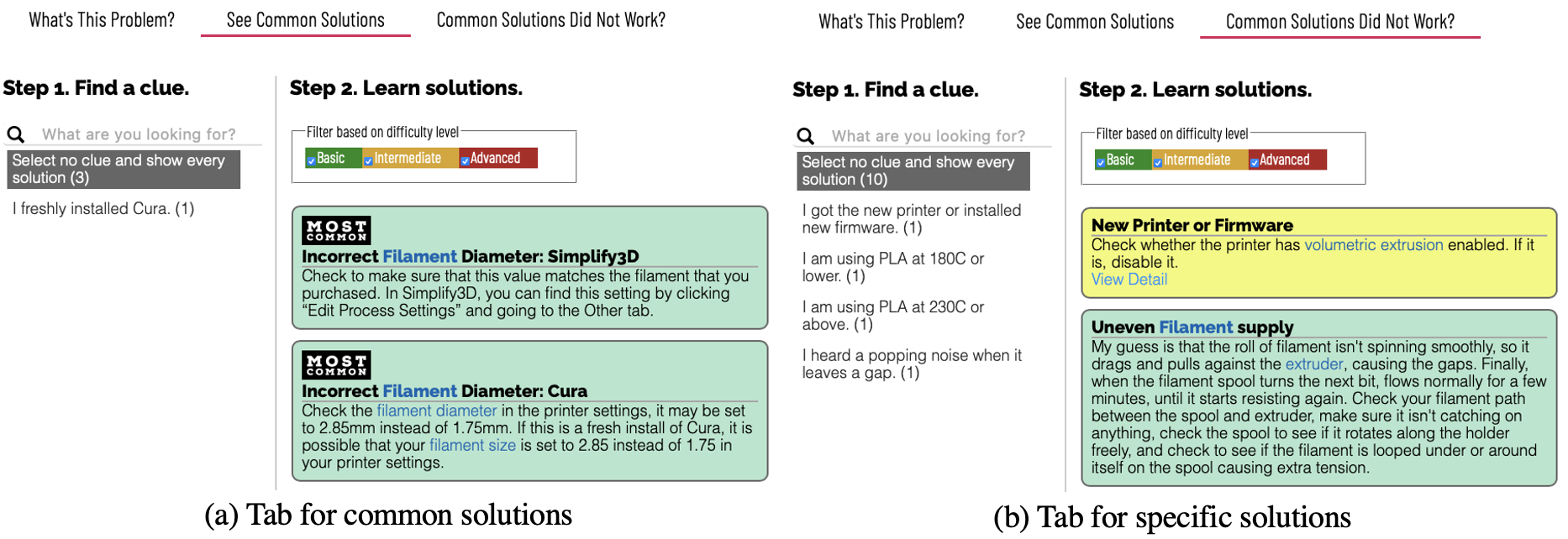}
    \caption{The second and the third tabs of the bottom section are for investigating solutions. Users usually start with ``See Common Solutions'' where they can either search for existing clues in ``Step 1. Find a clue.'' (on the left), or go straight to ``Step 2. Learn solutions.'' (on the right) to browse the solution cards they think are relevant to their 3D printing failures. Solution cards are color-coded by their difficulty levels (Basic is green, Intermediate is in yellow, and Advanced is in red). A difficulty-level filter is also available. If users cannot find the solutions they need in the common solution tab, they can go to the last tab to see a more comprehensive batch of solutions provided by the system. All the navigation and filtering features are the same as in the previous tab.}
    \Description{}
    \label{figure:system_solutions}
\end{figure*}

\textbf{\#4. Curated solution searching flow (DR4):}
\system curates two types of solutions which leads users to follow a reasonable flow of troubleshooting to first look into the most general solution and then proceed to the custom or advanced solutions rooted in the visual information-seeking mantra~\cite{shneiderman2003eyes}.
Users can begin by exploring common solutions that are likely to solve many general issues by `See Common Solutions' (Figure~\ref{figure:system_solutions}, (a)).
They can also refer to the `clue' to have a deeper understanding of what other phenomena could be there, which will help them self-diagnose issues.
The clues about observations (e.g., popping noise while printing, the bed is wobbling) or settings (e.g., printing temperature was lower than 180 degrees when using PLA, installing new firmware) 
may lead users to the specific solution set.
For example, if a user has an under-extrusion with the PLA as a material with a temperature set lower than 180 degrees, a low temperature of the print nozzle is likely to be a cause of not being able to completely melt and fuse PLA to safely 3D print the material \cite{printing_temperature}. 
The first-class trial must be increasing the nozzle temperature to around 210 degrees which tends to work for many across various types of machines.
If the user can spot any potential cause from the list of \textit{clues}, the user can filter the relevant solutions by navigating through the given options.
The `Common Solution Did Not Work?' tab (Figure~\ref{figure:system_solutions}, (b)) unfolds solutions for more special cases, for those who were not able to fix issues by the most common solutions to try these advanced solutions.
They can also filter and sort solutions by difficulty level, where the color scheme of the solution cards reflects the difficulty level for intuitive comprehension.
While novices would like to try the easiest options first, `hard' solutions may alert them that they need some more devotion.
The solution card shows the title and a short description to convey the key information. 
If a user
needs more details such as how to deal with the Cura, one of the most widely used slicing software, to increase the nozzle temperature, they can click on the `View Detail' or `View Video' in the card that will redirect them to the relevant online web document or a how-to video if needed.







\subsection{Example User Scenario}
Here we present an example user scenario of \system with a speculative user Teddy.
After auditing a Maker workshop from a local library and being introduced to many affordable 3D printers, Teddy recently bought a cheap 3D printer that costs less than \$100 for his hobby projects at home. 
Being unable to meet with his workshop instructors due to many canceled in-person sessions with the pandemic, he followed many how-to videos to install and set up the printer, then tried his first 3D printing using a 3D file he downloaded from Thingiverse~\cite{thingiverse}.
Yet, he found that his 3D print has a lot of gaps on the surface, which makes the outer walls look separated from each other. 
While he is eager to troubleshoot by himself, lots of general tutorials in how-to videos do not show how to address this specific issue. 
In fact, Teddy does not know what to call it nor how to describe the issue clearly, making the text search nearly impossible.
Reminding of many supporting communities learned from the workshop he attended, he posted a photo of it to seek help from community users. 
Unfortunately, his post did not get any attention.

He instead decided to use the \system.
In a few seconds, he got his issue diagnosed, learning that it is a known issue called ``under-extrusion'', and might be because of too low temperature set for printing or potentially humid filaments, and many more. 
As the system shows grayscale saliency maps that highlight the area with the unique visual characteristics as shown in Figure~\ref{figure:system_diagnosis}, he is also able to learn about possible different failures that he may need to take another look for his future print trials. 
Although his issue seemed to be obviously under-extrusion seeing the visual similarity between his and the sample image, 
Teddy also wants to check all other possible failure types to make sure that 
there are no other related issues. 
He taps on the `Show All Failure Types' button and checks the sample images informing common visual characteristics, under the `What's this Problem?' as shown in Figure~\ref{figure:system_diagnosis}.
Looking at all possible options, Teddy is confident that it is under-extrusion indeed, and learns how he can identify the same issue appearing in various forms in the future. 
Now he becomes more knowledgeable in that he needs to check whether there are any `spongy patterns' on the surface.

Stepping to the solution stage, Teddy tried all the common solutions suggested for under-extrusion, but his issue remains unsolved. 
He then proceeded to the specific solutions tab which covers most possible cases.
By going through the \textit{clues} that further hint about settings that cause the issue, he found `I heard a popping noise when it leaves a gap' matches his current settings. 
He expands the card to see relevant solutions, which noted that it might be caused by wet filaments.
He thoroughly reads the linked web article, learning that it can even exacerbate his printing quality by other akin problems such as blobbing.
One recommendation was to dry it using special equipment, while he chose to buy a new filament that was delivered in a sealed vinyl and cleared the issue.

\subsection{Implementation} \label{technitcal_implementation}
The back-end framework of the web-based user interface is built on Flask \cite{Flask}, a lightweight web application framework written in Python. 
Having a Python-based back end enables real-time model classification and visual explanation generation using our pre-trained deep learning models, post-hoc explanation techniques (Grad-CAM), and data visualization libraries used (PyTorch, Matplotlib, OpenCV). 
The front end is written in basic HTML, CSS, JavaScript, and additional libraries (D3, jQuery) for dynamic elements.
The visual explanation is generated by Grad-CAM \cite{selvaraju2017grad} PyTorch library, which is a popular post-hoc technique that computes the gradient saliency maps~\cite{simonyan2013deep}
to visualize the model prediction by highlighting ‘important’ pixels (i.e., changes in intensities of these pixels have the most impact on the prediction score).
All maps are transformed into grayscale by applying a sigmoid (slope = 4) to the model-generated attention scores pixel-by-pixel. 
This allows segmenting the attention areas (transparent mapping) from non-attention areas (dark mapping). 
Users can assume that a model's judgment of a failure type was made by looking at the attention areas of this saliency map. 
The brighter the mapping, the higher the attention of the model. Then, each saliency map is passed to the front end as a clickable tab for each prediction. 
By choosing one, the interface will show the corresponding introduction, clues, and solutions relevant to the selected failure type at the bottom for the user's further investigation.


\subsection{3DPFIX Dataset}

At a high-level, our system was implemented by extracting and organizing domain knowledge from ample Q\&A records stacked in an online community for 3D printing troubleshooting discussions.

\begin{enumerate}[leftmargin=.3in]
    \item To perform automatic 3D printing failure diagnosis:
    \begin{itemize}
        \item We collected a raw post dataset that contains user-uploaded failure images and corresponding text discussions.
        We leveraged the rich text information that is associated with user-uploaded photos in Reddit posts, including body texts and discourses in threads, and developed a keyword-based automatic classification of the posts and thus the images.
        \item Using this labeled image dataset, we trained individual binary classification models that manage one failure each. 
        The resulting classifiers can be applied to detect 3D printing failures from the user-uploaded photo.
    \end{itemize}
    \item To suggest solutions based on the diagnosed problems:
    \begin{itemize}
        \item We used the classified posts to extract encyclopedic user suggestions addressing each 3D printing failure. 
        \item We then extracted high-quality comments by the user feedback score (upvotes) along with \textit{clues} that indicate the specific cases from relevant threads as the most viable solutions.
    \end{itemize}
\end{enumerate}



\subsubsection{Data Preparation}
We collected a multi-modal dataset from FixMyPrint subreddit with 28,030 images from raw 30,780 posts. This dataset covers all posts of this subreddit during its existence from 2014 to July 2021.
The posts follow a common thread structure, with one post and subsequent comments. Often, multiple users discuss a specific 3D printing issue in varying depth in one thread.





\subsubsection{Building Image Dataset by Text-based Classification}

In order to automate the image-based diagnosis of failures through AI and to build a solution set that covers various cases accumulated over time in the community, we built an image dataset annotated with corresponding 3D printing failures. 
FixMyPrint's multi-modal (image and text) post dataset can work as a good source to determine the 3D printing failure type that each thread is trying to address. This grants us a unique opportunity in using the conversations to automatically label the images in the posts.
Thus, our initial goal became developing a text-based automated classification approach to categorize the 3D printing failure images of each thread.
First of all, as the discussions in this subreddit are highly domain-specific, they are different from everyday conversations or news articles that large language models are often trained on such as BERT \cite{huggingace_pretrained, huggingface, wolf2020transformers}. 
We decided to introduce heuristics into the classification and explore keywords-based approaches, as zero-shot classification techniques \cite{wolf2020transformers} seem not to be a viable solution owing to their low accuracy in the classification of posts. 
We used existing well-known troubleshooting archives as base documents to be compared with the posts, as those resources such as Simplify3D provide a pre-defined list of common 3D printing failures. 
For each printing failure type, they also provide an accompanying document with a detailed description of the phenomenon, why it may happen, and potential solutions for users. 
We specifically chose Simplify3D's print quality guide~\cite{simplify3d}, which presents 27 types of failure, so started with 27 base documents.
We conducted an initial automatic classification of posts using cosine similarity \cite{alodadi2015similarity} of the base documents on 26,232 posts that have at least one comment.
Also, to increase the accuracy of keyword-based classification without any overlapping keywords shared in different document themes, we defined one representative word for each failure type, which we call `failure-specific keyword', that can symbolize a unique failure type.
The leading researcher manually weighted the failure-specific keywords for each type, (e.g., not sticking to the printing bed - bed, stringing or oozing - string, layer shifting - shift, under-extrusion - extrude, warping - warp, blobs - seam, poor surface above supports - support).
We chose 5 of the most common types with distinct visual features as a starter to build an initial image dataset for a system demonstration.
Since this work is not a deployment study, we chose to explore a small set of failure types first, then plan to expand the coverage by investigating more general ways to reduce manual labor. 
Increasing diversity of printer machines and advanced materials would alter the dimensions and complexity of un/known printing failures being discussed in the community in the future, we are to present a pipeline that further facilitates human-AI collaboration to quickly adapt to new trends in 3D printing. We detail this process in Section~\ref{section:pipeline}.
Upon agreement between two researchers including one 3D printing expert, we combined the similarity score results and validated the results with the high similarity using the failure-specific keywords (e.g., stringing - string, warping - warp).
This approach achieved 73\% accuracy from 100 randomly sampled posts from the whole dataset, for the selected five types indicating a decent performance of the keyword-based classification.
. 

\subsubsection{Human Expert Validation on Image Data \& Technical Evaluation of Classification Models}

\begin{figure*}[t]
    \centering
    \includegraphics[width=1\linewidth]{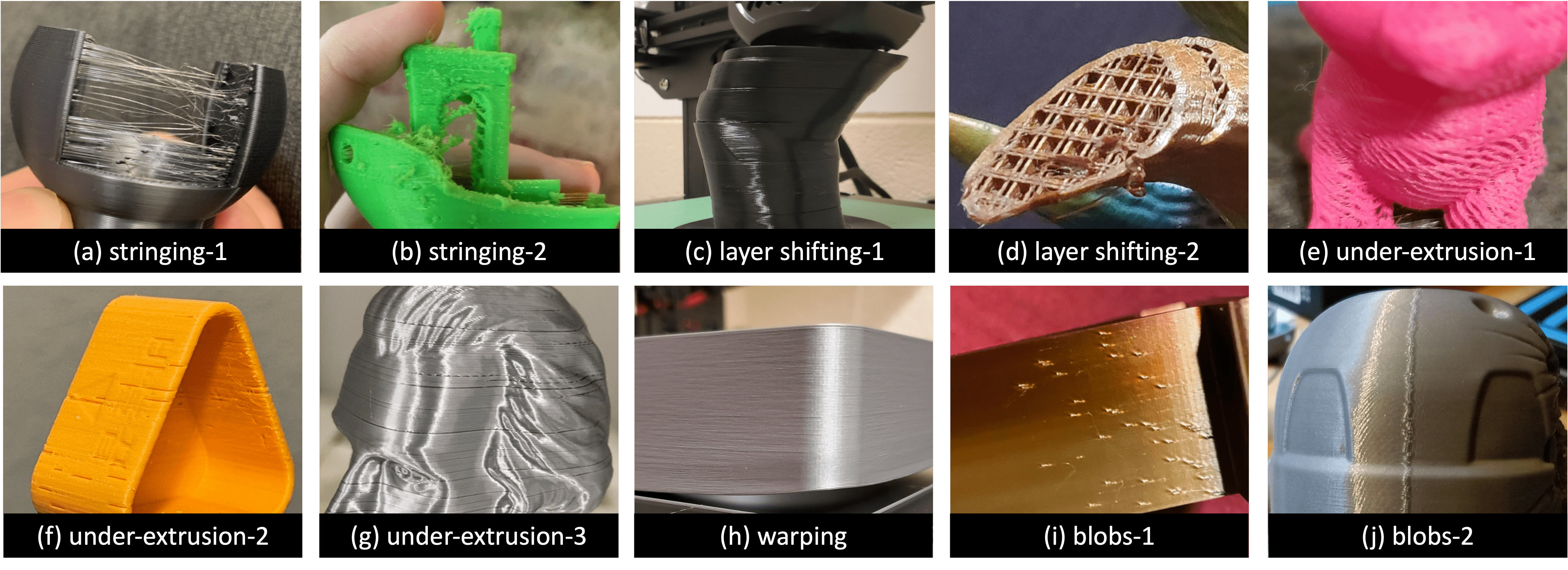}
    \vspace{-0.7cm}
    \caption{Distinct visual features of five selected failure types: (a) stringing - fine strings, (b) stringing - branches, (c) layer shifting - shifting in a side view, (d) layer shifting - shifting in a top view, (e) under-extrusion - spongy surface, (f) under-extrusion - empty gaps on the surface, (g) under-extrusion - missing layers, (h) warping - bending at the bottom of the model, (i) blobs - sporadic zits, (j) blobs - z seam.}
    \Description{There are ten example images (five images at the top row and the other five images at the second row). (a) image has fine strings in the middle of two object parts. (b) image has a benchy object with a branch-like string in the middle. (c) image shows a black 3D object that has sequential shifting on the top. (d) image shows a rainbow-colored 3D object that has shifting at the top view. (e) image shows a pink 3D object that has a distinctive spongy pattern on the lower surface. (f) image shows an orange 3D object that has slight and short gaps on the surface. (g) image has longer empty gaps for the multiple layers. (h) image shows a gray 3D object that has a bending at the bottom which is right on the printing bed.}
    \label{figure:visual_feature}
    \vspace{-0.5cm}
\end{figure*}

After the automatic classification using images, a human expert validated the classification results, defining the labels. 
In this process, we empirically found that there could be multiple visual characteristics in each failure type that we can categorize, as some examples in Figure~\ref{figure:visual_feature} show.
This finding was used to inform users of the common visual characteristics (as in Figure~\ref{figure:system_description}). 


\begin{figure*}[!b]
    \centering
    \includegraphics[width=0.8\linewidth]{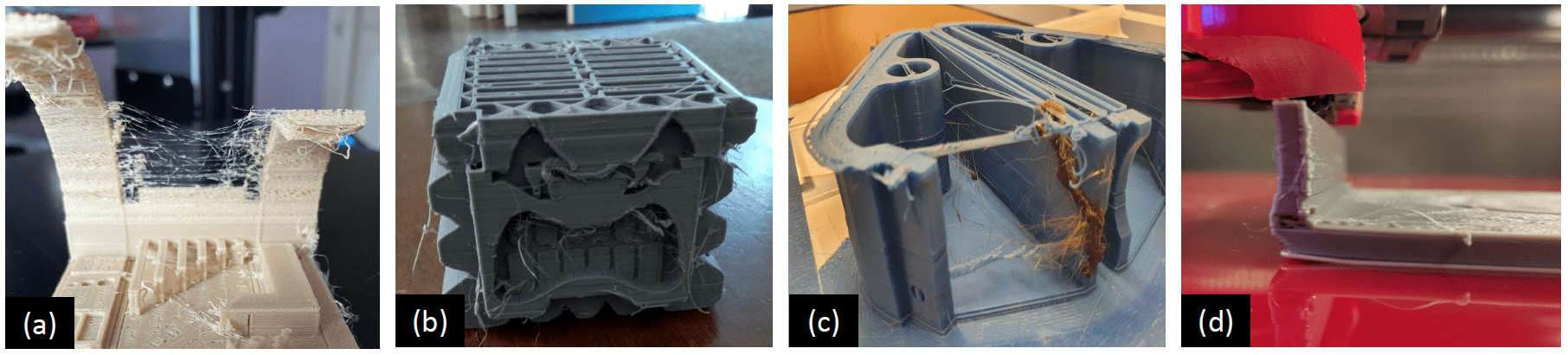}
    \vspace{-0.3cm}
    \caption{Example images of prints that have multiple failures from the FixMyPrint dataset: (a) stringing and under-extrusion, (b) layer shifting and stringing, (c) layer shifting and stringing, and (d) layer shifting and warping}
    \Description{There are four images aligned horizontally. The first image shows a white 3D object for calibration which has stringing in the middle and under-extrusion on the surface. The second image shows an object of a cube shape and has a layer shifting in the middle and strings at the lower part. The third and fourth image shows the gray 3D object.}
    \label{figure:multiple_failure}
    \vspace{-0.3cm}
\end{figure*}

The human expert validation finalized the training dataset for stringing (\textit{N}=439), layer shifting (\textit{N}=631), under-extrusion (\textit{N}=283), warping (\textit{N}=199), and blobs (\textit{N}=226), respectively. 
As real-world images in our dataset in Figure~\ref{figure:multiple_failure} shows, some prints might have visual characteristics of multiple failures at the same time.
For example, a typical test 3D model for calibration in Figure~\ref{figure:multiple_failure} (a) has visual features of stringing (fine strings, branch-like structure) along with a spongy pattern on the surface, which implies under-extrusion.
On the other hand, Figure~\ref{figure:multiple_failure} (b) and (c) are showing stringing and layer shifting, but with different visual features of it.
To capture these multiple classes, we trained a binary classification model (ResNet18 \cite{resnet}) that one model can determine whether the input photo contains one type of failure.
In this manner, one model is responsible for extracting the distinctive features of one failure type at a time.
To avoid confusion during training, we also set the negative class samples for the training set that do not present any distinct visual features.

\setlength\arrayrulewidth{1pt}
\begin{table}[]
\begin{tabular}{l|cc|cc|cc}
\hline
\multirow{2}{*}{\textbf{Failure type}} & \multicolumn{2}{c|}{\textbf{Training set}} & \multicolumn{2}{c|}{\textbf{Validation set}} & \multicolumn{2}{c}{\textbf{Test set}} \\ \cline{2-7} 
                              & ACC                  & AUC                 & ACC                   & AUC                  & ACC                & AUC               \\ \hline
Stringing                     & 97.15                & 99.66               & 88.51                 & 92.59                & 81.89              & 96.13             \\
Under-extrusion               & 95.55                & 99.18               & 82.14                 & 87.89                & 85.88              & 93.55             \\
Layer shifting                & 94.19                & 98.27               & 77.78                 & 81.69                & 73.61              & 81.10             \\
Warping                       & 94.12                & 98.51               & 87.18                 & 90.21                & 80.99              & 89.37             \\
Blobs                         & 95.57                & 99.26               & 91.11                 & 95.17                & 87.50              & 92.03             \\ \hline
\end{tabular}
\caption{ACC (Accuracy) and AUC (Area under the ROC Curve) of binary classification models for 5 types of printing failure within 20 epoch with the batch size of 32. All models were trained using ResNet18, Adam optimizer, and 0.0001 as a learning rate. The models were initialized with the pretrained weights, and we fine-tuned entire layers.}
\Description{Table that shows the prediction accuracy and AUC for training, validation, and test set. Test accuracy is 81.89\% for stringing, 85.88\% for under-extrusion, 73.61\% for layer shifting, and 80.99\% for warping. AUC is 96.13\% for stringing, 93.55\% for under-extrusion, 81.10\% for layer shifting, 89.37\% for warping, and 92.03\% for blobs}
\label{table:model}
\vspace{-10mm}
\end{table}

Since we carefully chose the training examples that present distinct visual cues for failures as shown in Figure~\ref{figure:visual_feature}, our current dataset has a relatively small number of samples. We still chose a CNN-based approach to automate the classification over more classic algorithms (e.g., SIFT) or traditional ML approaches (e.g., XGBoost, random forest). This is because our dataset is scalable as more cases can be discovered by communities. Also, even the same failure type can show different visual features. We believe that using CNN-based approaches can effectively address not only the large dataset but also automatically extract visual features without the hassle of defining manual features as the dataset grows.
We chose ResNet18~\cite{resnet} model for training and used Adam optimizer~\cite{kingma2014adam} with a 0.0001 learning rate. 
We initialized weights with the pretrained ResNet18 model
and fine-tuned all layers during training.
All five binary classification models converged within 20 epochs with a batch size of 32.
Four models for stringing, under-extrusion, warping, and blobs showed an 80+\% accuracy for the test set, while the layer shifting model showed slightly less test accuracy of 73.61\% as summarized in Table~\ref{table:model}.
All five models showed a fairly good Area Under the Curve (AUC, \cite{ling2003auc}) with at least 80\% AUC for the test set, implying that our models perform well in detecting one failure type overall.

\subsubsection{Constructing the List of Suggestions by Extracting Key Conversations} 
By combining the automated text-based classification and the image-based human expert validation, we finally got the post dataset annotated with the failure types.
Aiming to extract various quality suggestions from labeled community posts per each failure type, this approach ensures the freedom to choose from multiple viable solutions. 
As there could be multiple causes and suggestions for a single problem, 
it is critical to cover a wide spectrum of failure cases accumulated in the community. 
To create a thorough list of suggestions stacked in the community discussion, we first retrieved all comments that belong to each failure type and sorted comments by the number of upvotes (positive feedback) that the comments got from other users. 
Here we chose the upvotes count as it can reflect their quality, how helpful the contained suggestions are.
One researcher went through the comments sorted in descending order of the upvote count per each failure type and extracted the list of suggestions from the comments.
In this process, we extracted \textit{clues} from the suggestions (See Figure~\ref{figure:system_solutions}) which can specify the causes of failures and thus can lead to the specific solution set. 
We then used an online archive document~\cite{simplify3d} again to see if the suggestions are already included in the archive (\textit{common solutions}). If not, we defined those suggestions as \textit{specific} solutions.
As described in Figure~\ref{figure:system_solutions}, this separation between common \& specific solutions was used to guide users to the reasonable flow of troubleshooting, as common solutions can resolve the majority of issues.

\section{S2: Summative Study}





We evaluate \system through two steps. First, we conducted an experimental study that lets participants complete the predefined tasks in the controlled lab setting. We aimed to understand how several sub-components of \system driven by our design requirements can individually work in improving remote novices' current practice in their troubleshooting process.
To compensate for the artificial settings of the lab study and obtain more real-world feedback in natural and situated settings, we deployed \system for 3 weeks and encouraged active 3D printing users to freely use \system with the photos of their own failures that they have experienced so far.

\subsection{Methodology}

\subsubsection{Study Structure}
We first conducted an experimental study. Notable heuristic metrics (e.g., NASA TLX~\cite{hart1988development} or System Usability Scale~\cite{bangor2008empirical}) have been developed and widely used to measure the general usability of an interactive system as a whole.
However, complex interactive systems have many components (features) that support different objectives for each, thus adopting such heuristic methods can hamper researchers from understanding how each sub-component helps users achieve a certain goal in detail~\cite{hong2020towards}.
Similarly, 3D printing troubleshooting can be done by collectively comprehending various elements, such as recognizing visual characteristics of a failed print, learning what can cause the specific problem, and finding the most applicable solutions to try out; \system consists of several features to support users to achieve those elements.
We aim to measure how \system's components can improve users' experience in achieving the following objectives: (1) diagnosing/understanding failure types, (2) finding the applicable solutions, and (3) learning 3D printing knowledge and technical terms.
We extracted these 3 main metrics by reviewing 
Besides, HCI researchers have gradually accommodated applying the experimental study as a method to evaluate an interactive system (e.g., ~\cite{chaun2023FlatMagic}) under the two premises: (1) setting a baseline condition that can naturally capture a user's current practice, and (2) defining a task that is practical, specific, while having a clear ``ground truths'' to scientifically measure human users' task efficiency and task effectiveness.
Thus, we chose to do an experimental study for evaluation and followed this comparative structure to better reveal \textit{how each component of the system contributes to letting users fulfill the core sub-goals which will lead to successful troubleshooting}.
In our S2 setting, we define the baseline condition as participants' ``best practice'' using any online resources participants can use, following the most natural condition they can be under.
To prevent any bias, we did not specify the resources they can use in the study but allowed them to use any methods, mentioning some examples such as Google search, online articles, and online communities as examples. 
In our dry-run, we found out that given a specific online resource, subjects' behavior was highly bound to the certain resource in completing the tasks. We saw this is artificial and leads to biased action.
For the experimental condition, participants only used \system to do the same task without any other online resources.
We chose a within-subject study where participants were asked to perform troubleshooting using their familiar online resources (baseline condition: \textbf{C1}, hereinafter) and using \system{} (experimental condition: \textbf{C2}, hereinafter) for a direct comparison~\cite{lazar2017research}.



\subsubsection{Participants}
For our experimental study, in order to ensure the reliability of the study, we first decided to recruit at least 12 participants, which is the most common sample size used by CHI community~\cite{caine2016local}.
For recruitment, we used convenience and snowball sampling strategies~\cite{creswell2016}.
We first approached our acquaintances (1) who are interested in 3D printing or (2) working on a maker space in one of the two major research universities in the United States. 
We invited these contacts to participate in our study if they are interested in learning 3D printing but have no advanced knowledge about the domain. 
We also asked them to suggest potential participants who might fit into our inclusion criteria. 
Participation was voluntary and no compensation has been provided.
The IRB has been approved by the researchers' institutional boards at the moment we send an invitation. 
In total, 13 participants met our recruiting criteria (Male=5, Female=8).
Their age range between 18 and 34.
In the demographic survey, P8 and P10 self-evaluated that they have some extent of 3D printing knowledge and have experience in fixing troubles while the rest (\textit{N}=11) assessed themselves as novices---just started learning or have limited experience on troubleshooting.

\subsubsection{Study Procedure}






\begin{figure*}[!t]
    \centering
    \includegraphics[width=\linewidth]{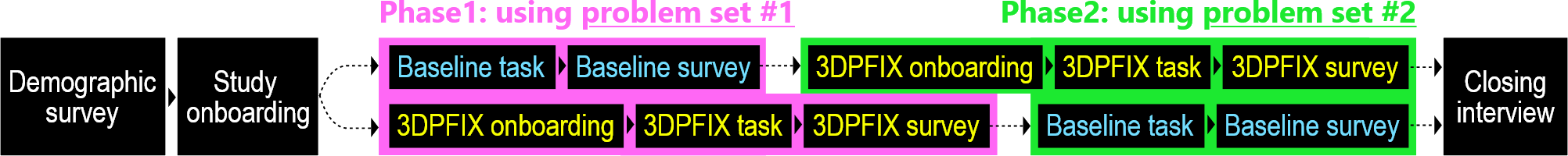}
    \vspace{-0.5cm}
    \caption{The flow of S2 experimental study.}
    \Description{}
    \label{figure:S2_flow}
    \vspace{-0.3cm}
\end{figure*}

Upon agreeing to participate, all participants were asked to finish the demographic survey. 
After the demographic survey, participants received a link to a remote, synchronous study link.
The lead author (the facilitator, hereinafter) met each participant one-on-one on Zoom to run the sessions and communicate with the last author after finishing each session.
At the beginning of the study, the facilitator asked for the e-signing of the consent form.
After consenting, the study began with the study onboarding session containing an introduction, the purpose of the study, and the study steps.
After finishing the study onboarding, we split participants into two groups for counterbalancing the ordering effect of two conditions.
We chose this within-subject design to capture participants' perspectives about the direct comparison between C1 and C2.
The first group of 7 participants used C1, baseline first (see the upper row in Figure~\ref{figure:S2_flow}) while the second group of 6 participants started from C2, experimental condition (see the lower row in Figure~\ref{figure:S2_flow}).
In C1, participants were encouraged to use any preferred troubleshooting strategies, including reading online archives, watching tutorial videos, and searching posts or posting questions to online 3D printing communities.
For C2, participants were only allowed to use \system{}; other online resources were prohibited.

To run the within-subject study, we prepared two problem sets. Set \#1 had examples of under-extrusion and layer shifting (PS1, hereinafter), and set \#2 had blobs and stringing (PS2, hereinafter).
To make their troubleshooting tasks realistic, we retrieved the four examples from the Ender3~\cite{ender3_subreddit} and 3DPrinting~\cite{3dprinting_subreddit} subreddits with experts' consult in terms of (1) how likely the problem might occur and (2) the complexity of the problem to solve it to control the level of difficulties between PS1 and PS2\footnote{The images in the problem sets are not included in our dataset used to train classification models.}. 
In distributing PS1 and PS2 to each condition, participants always used P1 first (i.e., Phase 1 in Figure~\ref{figure:S2_flow}) and P2 next (i.e., Phase 2 in Figure~\ref{figure:S2_flow}), regardless of the order of conditions.
Thus, we methodologically decoupled the problem sets and conditions.

Upon finishing each condition, participants completed the survey designed to evaluate their perceived and behavioral task efficiency, task effectiveness, and learning.
Detailed questions are explained in the later subsection.
In the first part of \textit{task performance} survey, we collected their behavioral data and attitudinal perception about the condition they finished.
In the later part of the \textit{learnability performance} survey, we evaluated how each condition facilitated their learning of given failure types and helped them gain knowledge about 3D printing in general.
For instance, we showed two different images and asked participants to identify the failure types.
We note that the learnability survey was only asked in their first trial because participants can gain an understanding of four different types of failure in their first trial already affecting their performance on the following condition, depending on how much they are engaged in using and learning.
For those who already exhaustively browsed every possible failure type in the first condition, the knowledge might be used to answer the learnability survey in their second condition.

After finishing both conditions, all participants are invited to a semi-structured interview. The interviews went for 27 minutes on average. 
In the interviews, we focused on capturing the aspects of: (1) what are the notable benefits and drawbacks of using \system in 3D printing troubleshooting and how the features in the system are related, (2) personal experience based on a direct comparison between \system and their current practice, (3) possible improvements of \system and how new design can improve people's 3D printing troubleshooting experience in general.
\system was deployed as a web application (Available at [Anonymized for the review]) that participants can access using any modern web browser. Participants were asked to share the screen during the study while the whole process is recorded for further analysis.





\subsubsection{Measures}

\begin{table}[!t]
\begin{tabular}{l|ll|ll|l}
\hline
 &
  \multicolumn{2}{l|}{\textbf{Task efficiency}} &
  \multicolumn{2}{l|}{\textbf{Task effectiveness}} &
   \\ \cline{2-5}
\multirow{-2}{*} &
  \scriptsize{Failure identification} &
  \scriptsize{Solution-seeking} &
  \scriptsize{Failure identification} &
  \scriptsize{Solution-seeking} &
  \multirow{-2}{*}{\textbf{Learnability}} \\ \hline
\textbf{Behavioral} & Q1 & Q2 & Q1 & Q2 & Q8 \\
\textbf{Attitudinal} & Q3   & Q5   & Q4 & Q6 & Q7a--Q7d \\ \hline
\end{tabular}
\caption{Type of measures in S2 and mapping of corresponding questions in our survey design.}
\vspace{-0.8cm}
\label{table:S2_measures}
\end{table}

We measured participants' behavioral \& attitudinal \textit{task efficiency} (i.e., how using a system helped them cut down their effort), \textit{task effectiveness} (i.e., how using a system helped them to yield a quality outcome), and \textit{learnability} (i.e., how did using 
a system help extend their internal representation of the 3D printing domain). 
In measuring the three directions, we designed the survey to capture a user's behavioral performance and attitudes towards C1 or C2.
To measure task efficiency and task effectiveness, we broke down the measure into (1) failure identification: detection of an accurate type of 3D printing failure, and (2) solution-seeking: derivation of a reasonable action plan that can result in desirable outcomes. 
We did not ask participants to measure the time to complete each task not to increase their workload. Instead, the first author reviewed the video recordings of the study and calculated the time taken to measure behavioral task efficiency for failure identification and solution-seeking. 
Table~\ref{table:S2_measures} shows how our measures are mapped with the specific questions we implemented through our survey. 

The specific survey questions we implemented are as follows:
\begin{itemize}
    \item \textbf{Q1}. What was the type of failure you identified? Please explain why.
    \item \textbf{Q2}. After your search, what would you do to fix the type of failure? Explain why you thought your plan can work.
    \item \textbf{Q3}. I was able to identify \textbf{the type of failure} with less effort and time.
    \item \textbf{Q4} I was able to identify \textbf{the type of failure} accurately using the tool(s).
    \item \textbf{Q5}. I was able to find \textbf{the solution} with less effort and time.
    \item \textbf{Q6}. I was able to identify \textbf{the solution} that would accurately work for this type of failure.
    \item \textbf{Q7a}. I was able to learn about \textbf{the type of failures} that the given images show in this condition.
    \item \textbf{Q7b}. I was able to learn the \textbf{3D printing-related knowledge} about this 3D printing failure.
    \item \textbf{Q7c}. I was able to learn about \textbf{useful suggestions and tips} that could resolve failures.
    \item \textbf{Q7d}. I was able to learn about \textbf{logical flow of troubleshooting} (problem identification - finding a clue/situation indicates the issue - solution-seeking) in 3D printing.
    \item \textbf{Q8}. [Attach an unseen image] What is the type of failure this image shows?
    \item \textbf{Q9}. Are there any other thoughts that you would like to share about your experiences?
\end{itemize}

In defining questionnaires above, we referred to standard usability-based questionnaires (e.g., NASA TLX~\cite{hart1988development}, SUS~\cite{bangor2008empirical}) and similar prior works (e.g., \cite{chuan2022flatmagic, choi2019aila}) that evaluated their interactive systems through an experimental study. Based on the prior work, we adapted the questions to our three major metrics, mostly focusing on behavioral/attitudinal efficiency (time taken) and effectiveness (accuracy) for each metric. Additionally, for learnability, we added major obstacles identified (e.g., problem articulation, finding applicable solutions, difficult technical terms) to be components to measure. 
In implementing our survey, we made Q1, Q2, Q8, and Q9 open-ended short answers while the rest rated on a 1-7 Likert scale, ranging from strongly disagree to strongly agree.
To evaluate the quality of participants' answers in Q1 and Q2, we recruited one 3D printing expert who has been using 3D printers for 4.5 years and has extensive knowledge in 3D printing and experience in troubleshooting. 
All participants' answers and condition information were de-identified when provided to the 3D printing expert for analysis. 
Given the participants' responses including failure identification and the solution they chose to try, the expert rated participants' failure identification with True/False and the solution in a 1-7 Likert scale in terms of quality.
In analyzing Q1 and Q2's behavioral effectiveness, we used Pearson's Chi-squared test for null hypothesis analysis by calculating the frequency of True and False labels under two conditions. For all ordinal responses in 
Q1 and Q2's behavioral efficiency
and the 1-7 Likert Scale (Q3-7), we used the Kruskal-Wallis test.
We used Python packages such as Scipy and statsmodels for all statistical analyses.
Also, the facilitator used an iterative coding method~\cite{saldana2015coding} to analyze qualitative data. Upon completion of each interview, we first built codes to extract unique or meaningful aspects based on responses in the transcript and wrote an analysis down to combine insights from different interviews and synthesize them. After finishing all interviews, we finally did a consensus-based diagramming to structure what we found.

\subsubsection{Deployment Procedure}
The experimental study enabled us to understand how the several sub-components work together to help users troubleshoot the failures.
To see how \system can holistically work with users in more realistic, less-artificial settings, we deployed \system for a short time, to obtain feedback from active 3D printing users who had already experienced failures.
We distributed a flyer through the online communities, including subreddits (e.g., FixMyPrint~\cite{fixmyprint}, 3DPrinting~\cite{3dprinting_subreddit}, Ender3v2~\cite{ender3_subreddit}), discord servers (e.g., Creality 3D~\cite{creality_discord}), and online forums (e.g., 3D Printing Space~\cite{3dprintingspace}), that have an active discussion about 3D printing matters, such as troubleshooting, materials, and printers.
Users were encouraged to use the photos of the failed prints that they have experienced so far,
\IssueOneA{without any verbal/functional restrictions of failure types},
and submit voluntary feedback through Google Forms. Since it was fully voluntary, no compensation was provided to the users.
In the survey, they were asked to answer four optional questions:
\begin{itemize}
\item How would you describe your overall experience using 3DPFIX? How do you think 3DPFIX could be helpful for novices?
\item Do you find any difficulties or problems using 3DPFIX?
\item Would you recommend 3DPFIX to other 3D printing novices? Why or why not?
\item Do you have any additional comments to improve 3DPFIX?
\end{itemize}
From the deployment for 3 weeks, \system obtained 30 upload attempts and written feedback from 6 3D printing users (U1-6).



\subsection{Results}

\begin{table}[!b]
\begin{tabular}{c|l|cccc}
\hline
\textbf{Category}                                & \multicolumn{1}{c|}{\textbf{Measure}} & \textbf{N} & \textbf{p-value} & \textbf{\begin{tabular}[c]{@{}c@{}}Effect Size\\ (Cohen's d)\end{tabular}} & \textbf{\begin{tabular}[c]{@{}c@{}}Power\\ at N\end{tabular}} \\ \hline
\multirow{2}{*}{\textbf{Failure Identification}} & Task efficiency (Q3)                  & 13         & 0.000107         & -2.09471                                                                   & 1                                                             \\
                                                 & Task effectiveness (Q4)               & 13         & 0.000148         & -2.01261                                                                   & 1                                                             \\ \hline
\multirow{2}{*}{\textbf{Solution-seeking}}       & Task efficiency (Q5)                  & 13         & 5.86E-05         & -2.43937                                                                   & 1                                                             \\
                                                 & Task effectiveness (Q6)               & 13         & 0.000321         & -2.0381                                                                    & 1                                                             \\ \hline
\multirow{5}{*}{\textbf{Learnability}}           & Behavioral learnability (Q8)          & 7          & 0.012475         & -1.63979                                                                   & 1                                                             \\
                                                 & Failure identification (Q7a)          & 7          & 0.048668         & -1.30191                                                                   & 0.99                                                          \\
                                                 & Technical knowledge (Q7b)             & 7          & 0.095791         & -0.78662                                                                   & 0.74                                                          \\
                                                 & Troubleshooting tips (Q7c)            & 7          & 0.001879         & -3.17091                                                                   & 1                                                             \\
                                                 & Logical flow (Q8d)                    & 7          & 0.240433         & -0.69518                                                                   & 0.634                                                         \\ \hline
\end{tabular}
\caption{
Power analysis results of measures in S2 with the corresponding sample size. In evaluating learnability measures, only the first condition (\textit{N} = 7) was used due to a possible learning effect in the second condition.
}
\label{table:S2_power}
\end{table}

\begin{figure*}[!t]
    \centering
    \vspace{-0.0cm}
    \includegraphics[width=1.0\linewidth]{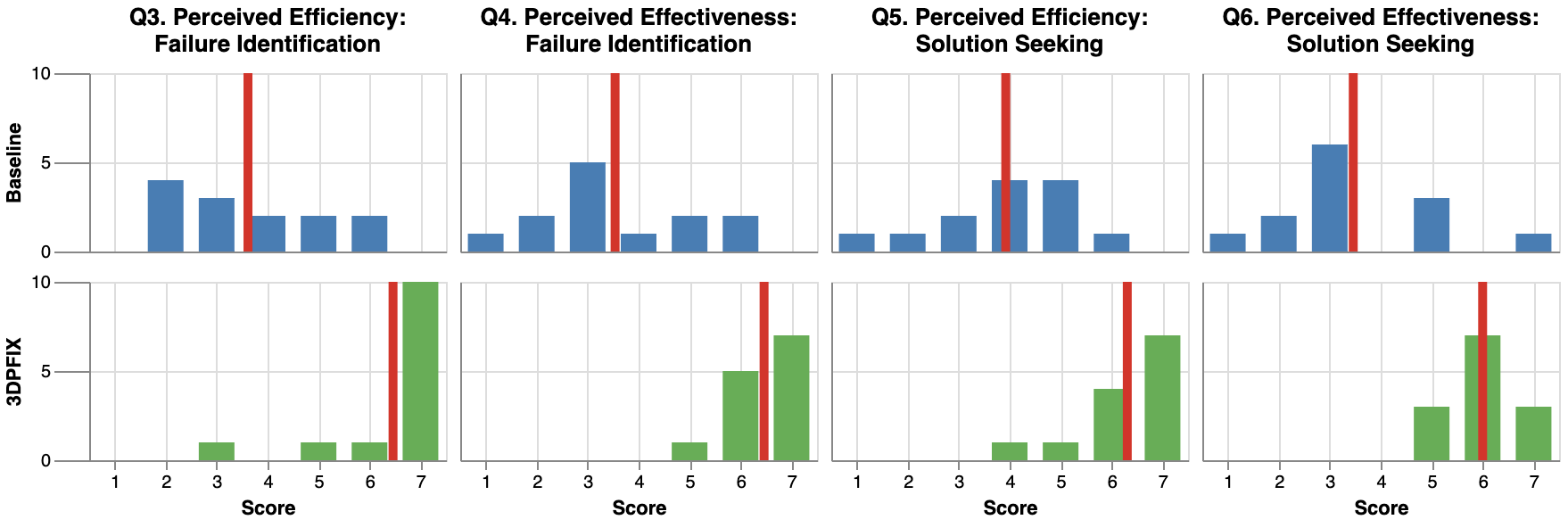}
    \vspace{-0.5cm}
    \caption{
    Distribution of 13 participants' responses
    about their perception on a 7-point Likert scale, where 1 represents `Strongly Disagree' and 7 is `Strongly Agree', under two different conditions: baseline using online resources (top, blue-colored bar charts) and experimental using \system (bottom, green-colored bar charts). A red line represents the mean value. For all 4 metrics on participants' perceived efficiency and effectiveness, using \system outperforms the baseline.}
    \Description{}
    \label{graph:S2_efficiency}
    \vspace{-0.0cm}
\end{figure*}

\begin{figure*}[!t]
    \centering
    \vspace{-0.5cm}
    \includegraphics[width=1.0\linewidth]{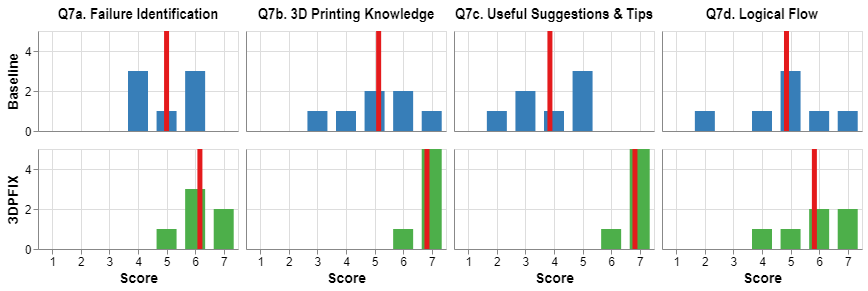}
    \caption{Distribution of participants' perceived learnability responses in two conditions: baseline using online resources (top, blue-colored bar charts) and experimental using \system (bottom, green-colored bar charts). When using \system, all 4 metrics (on a 7-point Likert scale: 1 is `Strongly Disagree' and 7 is `Strongly Agree') showed increased performance compared to the baseline.}
    \Description{}
    \label{graph:S2_learnability}
    \vspace{-0.5cm}
\end{figure*}

As shown in Figure~\ref{graph:S2_efficiency} and Figure~\ref{graph:S2_learnability}, \system outperforms the baseline. 
As the distribution does not follow the normal distribution, we used a Kruskal-Wallis test to compare the two groups' perceived efficiency (Q3, 5), perceived effectiveness (Q4, 6), and behavioral learnability (Q8). 
The power analysis results show that our data collected from 13 participants contains enough power (> 80\%) to correctly detect any significant difference between the baseline and experimental groups. The Cohen's d effect sizes are large enough so that a smaller sample size can still produce the desired amount of power for statistical testing.
Table~\ref{table:S2_power} shows power analysis results of all measures with 13 participants, including p-value, Cohen's d, and effect size.

\subsubsection{\textbf{Effective Problem Identification \#1: Finding the Right Representation}}
\system significantly reduced participants' effort in identifying and learning the accurate naming of failure types by simplifying the process of exploring the right representation (keyword seeking).
Using \system, their perceived efficiency on failure identification significantly improved than the baseline.
Compared to baseline ($M_{base}$ = 3.62, $SD_{base}$ = 1.44), participants thought that they were able to identify the failure type with less effort and time using \system ($M_{exp}$ = 6.46, $SD_{exp}$ = 1.15, \textit{p} < 0.005*). 
In terms of behavioral efficiency, there was no significant difference in the time taken to identify the problem (\textit{p} > 0.05), but there was about 50\% improvement on average completion time using \system ($M_{base}$ = 287, $M_{exp}$ = 253 seconds).
As a major difficulty of identifying the problem and finding solutions in the baseline, many participants pointed out finding the right representation for search (P6, ``Since I do not know the exact term, I started searching with the keyword, 3D printing problem'').
The difficulty of representation and the need for automated diagnosis were also recognized by users. \textit{``The system would especially be helpful to novices because they may not know the relevant keywords to search for issues and they may not even be aware of issues that are in their prints in the first place''} (U3).
During C1, we observed that all participants engaged in a loop for finding the proper representation, such as reading articles or other online archives, and picked the keywords from resources for another round of search (shift representation).
Most participants started from a general representation, such as `3D printing failures', `3D printer problems', and `3D printing common issues', since they do not know the exact technical terms to search. 
Some participants started with more descriptive representation by describing the visual characteristics in their own words. For example, P4 and P7 chose \textit{`3D printing dots, bubbles'} and \textit{`Bumps over the layer'} to describe blobs. P7 described the under-extrusion failure with \textit{`Outer shell is not sticking to inner mesh}'.
However, it was not always easy for people to articulate the failure, especially if the terms are more technology-oriented \textit{``For me, it was easy to search for stringing and shifting as the terms are more related to daily life---easy to describe with everyday words---, but for under-extrusion and blobs, it was not easy''} (P11).
\textit{``For some images, it is easier to describe, but for some situations, I don't know how to describe. In this case, it is hard to search, because I don't know the exact term [to begin with text-query for search]''} (P8).
Finding the right representation was still challenging for participants who have some printing experience (P10). 
P10 noted that she relied on her previous knowledge and experience. However, some failures are also new so she still needed an online search to investigate the right technical term.

\subsubsection{\textbf{Effective Problem Identification \#2: Understanding Visual Characteristics of Printing Failure}}
\system facilitated participants' effective problem identification by supporting them in understanding the common visual characteristics.
Using \system, their perceived effectiveness on failure identification significantly improved than the baseline.
Participants felt that the accuracy of failure identification significantly increased ($M_{base}$ = 3.92, $SD_{base}$ = 1.33, $M_{exp}$ = 6.31, $SD_{exp}$ = 0.91, \textit{p} < 0.005*).
Also, from the expert review of their answers, the behavioral effectiveness (Q1) (accuracy of the answers) using \system significantly improved than the baseline (\textit{p} = 0.006* by Pearson's Chi-square test).

No matter how descriptive their initial representation (search keywords) is, all participants used an image-oriented search strategy, searching similar images under the `Images' tab from Google search results or looking at representative images in the online articles.
P6 explained her approach started with the general representation and compared the images to locate similarities, and if it fails, she went through the image tab.
However, eight participants (P1, 3-6, 8, 11-12) noted their struggles due to 3D printing failures' multi-appearance feature. 
They found it difficult to obtain images with the same or similar visual characteristics.
Also, such processes demand a lot of time and effort.
\textit{``Images may not look the same as the ones in the online resources. It requires some knowledge to diagnose''} (P2).
This uncertainty eventually lowered the confidence in their decision.
P12 pointed out that she could not even find diverse images to compare through in her current practice. 

On the other hand, \system helped participants understand the visual features of printing failures and thus was granted confidence in the failure type detection from two key features: (1) grayscale saliency maps (See Figure~\ref{figure:system_description}) and (2) common visual characteristics of the failure (See Figure~\ref{figure:system_diagnosis}).
Twelve participants (P1, 3-13) emphasized their preference for the grayscale saliency maps (1), as it supports them to understand which part is problematic.
Highlighting the part to focus on, \system improved participants' confidence in the failure type that they identified by double-checking AI's decision on all failure types that \system supports.
\textit{``\system gives possible failure types, so I can take a look at all of those and pick the most relevant one with [confidence], I can make my own decision based on AI’s prediction''} (P3).
From the deployment, U1 mentioned: ``I really like that it shows examples of other possible failures. It helps a lot to compare and figure out the actual issue before troubleshooting. A lot of the time, I find myself troubleshooting for random possible issues without knowing the exact problem. This tool just really helps to [narrow down the choices and gives a great, clear explanation of each problem]''.
In addition, several participants mentioned that they relied on a grayscale saliency map in assessing the AI's trustworthiness. They mentioned having saliency maps helped them trust the model's results (\textit{N} = 5, P1, 4, 6, 8, 12), as focus given to unreasonable areas can help them to understand the AI made an unexpected mistake (P8). P1 remarked: \textit{``If mask---highlighted area---is not highlighting the print, the AI's decision is probably wrong. It is the way to double-check AI's decision''}.
However, one user (U3) from the deployment reported that the saliency maps were slightly hard to understand at first, which indicates that there should be additional descriptions about how to perceive the grayscale area of maps for laypeople.

Representative images that inform common visual characteristics of each failure also helped participants (P3, 4, 6-8, 12) identify the problem by comparing those representative images to the given images.
\textit{``That images with common visual characteristics really helped me find which problem is showing in the image. In the baseline,
I wanted to see other images with blobs but it took too much time to see if other people also have the problem that looks like the same as mine''} (P4).
Similarly, U1 also mentioned that \textit{``even if the AI's guess is wrong, there are multiple other examples for reference''}.
P6 particularly appreciated learning the specific terms to describe the common visual characteristics of the system (e.g., spongy pattern, gaps on the surface, and missing layers for under-extrusion in Figure~\ref{figure:system_description}), noting that \textit{``Only the experienced users can know [how to call it]. Since \system can inform the specific keyword for each visual characteristic, I can use this for the further search''}.

\subsubsection{\textbf{Effective Solution-seeking: Easy Guidance to Solutions in Depth}}
\system improved participants' ability to find quality solutions and to set their troubleshooting plan more easily, helping them acquire 3D printing knowledge.
Regarding the expert review of participants' solution quality, \system significantly outperformed baseline ($M_{exp}$ = 5.53, $SD_{exp}$ = 1.93, $M_{base}$ = 3.92, $SD_{base}$ = 2.32, \textit{p} = 0.006*).
In deciding what to do to resolve the failure that they identified (solution), the participants perceived that they were able to get solutions with significantly less effort and time (\textit{p} < 0.005*) using \system ($M_{exp}$ = 6.46, $SD_{exp}$ = 0.63) than the baseline ($M_{base}$ = 3.54, $SD_{base}$ = 1.50).
Even though correct answers were not given at the end, they felt more confident (\textit{p} < 0.005*) on the solutions they came up with using \system ($M_{exp}$ = 6.0, $SD_{exp}$ = 0.68) than the baseline ($M_{base}$ = 3.46, $SD_{base}$ = 1.55). Their confidence in their solutions under the baseline condition was significantly lower than that of using \system.
Interestingly, there was no significant difference in behavioral efficiency between the two conditions (\textit{p} > 0.05), and even participants spent more time finding solutions using \system on average ($M_{exp}$ = 500, $M_{base}$ = 376 seconds). 
Based on our observation, most participants showed more engagement when using \system. 
For example, they wished to explore every feature of the system and were more enthusiastic about learning by carefully going through the list of solutions and linked articles/videos. Even though they spent more time using \system than baseline, they perceived that they used significantly less time and effort in deciding solutions to try, which can indicate increased engagement with the less mental load we assume. 


Participants were able to effectively access a variety of solutions with depth using the \system. In particular, they mentioned that the following features of \system helped them to browse solutions: (1) providing common and specific solutions using different tabs, (2) `View Detail/View Video' enables participants to review further relevant information, (3) visualizing difficulty levels \& clue filtering, and (4) providing easy description when hovering over technical jargon (See Figure~\ref{figure:system_solutions}).
In the baseline, many participants suffered from too many online resources available whereas those were vague or too general. They do not know where to start and which suggestion is common and viable to start with.
Novice participants (\textit{N} = 11) were bewildered by the overflow of online resources as they found that many of them discussed similar suggestions in a very general manner. 
From the deployment, U5 raised the same concern by reflecting that searching on the internet can be overwhelming.
P3 complained about the resources she was able to find, mentioning that \textit{``It tells me to increase the printing temperature, [but how]?''}.
P8 was concerned about missing the most critical and useful information from online supporting communities, by scrolling and skimming through general but not quite reasonable solutions that she can trust.
On the contrary, participants (P4-5, 7-8, 12) appreciated the flow of \system, which leads users to read the common solutions first and more specific solutions if needed, by separating solutions with two different tabs, `Common Solutions' and `Common Solutions Did Not Work?'.
\textit{``People really get stressed when there is too much information at once, it is better to go step by step. For common solutions, I feel confident those are the things I was looking for''} (P3).
Providing direct links to relevant web resources for more detail using the `View Detail' and `View Video' buttons also supported participants' (P4-5, 7-9, 12) a better understanding of the solutions in-depth.
While the short description contained in each solution card (as shown in Figure~\ref{figure:system_solutions}) enables quick decision-making, the linked video further assists in understanding what to do: \textit{``For that suggestion about retraction distance, I could not understand what it does. That linked online resource did give me an answer to my question, `increase the distance, BUT what would it do?'''} (P9)
To novices, having the difficulty level labels and clue filtering function was helpful (P1-2, 4, 6-8), as it also gives them guidance on where to start \textit{``It gives where to start, I would start with the basic one''} (P2).
\textit{``I think I will start with the basic solutions and if I think that would work, I will check the clue. I will find the best one considering my situation''} (P6).
Participants also liked the easy navigation of \system. All participants mentioned that \system was easy to understand and how to use, and the simple interaction and layout were user-friendly. 

\subsubsection{\textbf{Improved Learnability: 3D Printing Knowledge \& Technical Terms}}

As Figure~\ref{graph:S2_learnability} shows, \system significantly improved participants' perceived learning abilities in three learnability metrics.
Participants perceived that they were able to learn about the failure types given as the task materials were significantly better (\textit{p} < 0.05). 
than the baseline ($M_{base}$ = 5.0, $SD_{base}$ = 0.93) when they use \system ($M_{exp}$ = 6.17, $SD_{exp}$ = 0.69).
Results proved that \system can improve participants' experience in getting 3D printing knowledge about specific failures ($M_{base}$ = 5.14, $SD_{base}$ = 1.25, $M_{exp}$ = 6.8, $SD_{exp}$ = 0.37, \textit{p} < 0.05).
Perceived learnability about troubleshooting tips was notably improved using \system ($M_{exp}$ = 6.83, $SD_{exp}$ = 0.37) than baseline ($M_{base}$ = 3.86, $SD_{base}$ = 1.12), which reflects that \system can provide more diverse \& exhaustive suggestions than the baseline (\textit{p} < 0.005*).
In understanding the logical troubleshooting flow, the two groups' responses did not significantly vary (\textit{p} = 0.24).
Still, we see the mean value ($M_{base}$ = 4.86, $SD_{base}$ = 1.46) was slightly higher by 18\% when using \system ($M_{exp}$ = 5.8, $SD_{exp}$ = 1.07). 
In acquiring technical terms, \system's hover-over function for the technical terms (as shown in Figure~\ref{figure:system_hoverover}) improved novice participants' ability to learn technical terms.
\textit{``If I am not using \system, I need to do another search for that term and get back to the sources that I was looking at. It takes twice the effort. The hover over really saved my time and effort''} (P4).
U1 also reported that the highlighted definition of common 3D printing terminology is specifically good for newcomers.

\subsubsection{\textbf{User Feedback \& Limitations}}
\IssueOneA{Our deployment flyer informed that \system is an ongoing project with plans for future expansion of its scope. Neither the flyer nor the system itself imposed any verbal or functional restrictions on users regarding the types of failure cases they could upload, ensuring a fair and natural setting. Despite the current scope of \system,} the majority of users (\textit{N} = 5) from the deployment favored the current ability and potential of \system for novices. 
They expressed their willingness to recommend \system to novices as a quick and efficient means of diagnosing problems (U1-4), obtaining immediate feedback (U2, 5), and learning about common failures (U1, 5) through an easy-to-follow workflow (U4). 
U6, while more neutral, highlighted the need for \system to expand its failure case knowledge base, which is one of the main focuses of our future work for a longitudinal deployment study.
As potential areas for improvement, two participants (U2, 5) shared their experiences with false positive issues. They mentioned that \system correctly identified the failure but also erroneously labeled a different type as `Highly likely' failure. 
U2 suspected that the photo orientation might be a factor, while U5 attributed it to a cluttered background. \IssueOneA{Despite maintaining anonymity by not collecting personal identity information, we observed that at least one participant attempted multiple uploads. 
U5, for instance, reported conducting two separate trials using photos of the same print, each taken against a different background. 
In their initial attempt with a cluttered background, the system incorrectly identified the background as an 'under-extrusion' issue. However, in the second attempt with a clear background, U5 noted that the system accurately diagnosed the real issue as 'blobs' without any false positives in the background.}
Both U2 and U5 suggested including clear instructions for novices on how to capture photos to achieve optimal AI performance. 
We believe that enhancing the classification model's performance, along with providing clear instructions, will be a primary focus of our future work to make \system more effective in real-world scenarios. 
Additionally, U3 pointed out that the solution tab is text-heavy and recommended incorporating visual aids to help users understand how specific methods can influence print results.
\IssueOneA{Lastly, within the first three weeks of deployment, we received 6 written feedback responses out of 30 upload attempts, recognizing the possibility that some users may have made multiple attempts. 
We encouraged users to voluntarily submit separate Google Forms following their interaction with the system. 
We acknowledge that the process of providing individual written feedback might have been daunting for some users, potentially resulting in a lower number of feedback responses compared to the total number of upload attempts.
In our longitudinal deployment study, we plan to implement a built-in feedback system that enables users to submit their feedback directly within the \system interface, rather than redirecting them to Google Forms. Additionally, we intend to expand our reach to potential user pools by distributing the flyer to makerspaces, printing centers, and 3D printing workshops. This broader outreach will allow us to gather feedback from users with diverse levels of expertise, enhancing our understanding of \system's performance and usability.}

\section{Discussion \& Implications for Design}

We offer high-level insights we learned through S1 and S2 in terms of the factors that system designers can consider when leveraging social annotation in building new applications for troubleshooting and a broader scope of applications for exploratory tasks. 

\subsection{Leveraging Social Annotation in Designing Troubleshooting Interfaces}
In applying social annotation in the domain of 3D printing troubleshooting, our main considerations were (1) considering a user-perceived complexity of sub-tasks, i.e., among several small and large interactions a user makes in troubleshooting, which sub-task typically blocks them from making progress, and (2) Social annotation-AI transfer feasibility; i.e., would using social annotation can result in building an AI model that performs well.
Literature in human-AI collaboration has shown that users would have a higher valuation of the AI-driven features if the task automated by AI is perceived cognitively challenging~\cite{frich2019mapping, chuan2022flatmagic}.
However, the way to realize AI-driven design and its performance would be bounded by the quantity and quality of the data. 
When defining the scope of the AI-driven features in supporting troubleshooting, designers may consider the following four segments:

\begin{itemize}
    \item \textbf{S1. Demanding sub-task, High-performing AI}: The segment of \textit{Promise}. The AI-driven feature can help novices through cognitive offloading, thus the feature in this segment would likely be perceived as useful. Designers may put the highest priority on building the AI that belongs to this segment.
    \item \textbf{S2. Demanding sub-task, Low-performing AI}: The segment of \textit{Challenge}. A designer can expect a highly positive user-side effect by implementing this feature. To realize this feature, designers may focus on diagnosing the factors that negatively affect performance.
    \item \textbf{S3. Easy sub-task, High-performing AI}: The segment of \textit{Potential}. While it is possible to build a viable AI-driven feature that works in a user's task, the way to offer the feature should be carefully considered and designed. Even if the task itself is not taxing, this feature can be useful if novices conduct this type of task repetitively throughout. For instance, for users who frequently ``translate'' unknown terms in a complex article, having a high-performing domain-specific lexical simplification model can be useful.
    \item \textbf{S4. Easy sub-task, Low-performing AI}: The segment of \textit{Unwanted}, no strong reason to implement the feature that falls into this segment, as user-side merit is unclear while the cost for developing the high-performing AI is expensive.
\end{itemize}

Considering the two factors could help a designer scope the AI's role in troubleshooting and reduce the risk of building an AI that would be perceived as not useful or not feasible. 
In our case, we found formulating a lexical query can be a learning block that hampers novices from building up their knowledge. 
Replacing the lexical query with an image-based approach was feasible due to the social annotation collected in online resources we targeted to use. 
In general, we think it's worth considering ranking the possible sub-tasks in terms of task complexity and feasibility considering available social annotation.

\subsection{Improving the Way to Communicate in Future Troubleshooting Designs}

Among the many learning blocks, the most evident block for the remote novices we identified in S1 was formulating a query using accurate domain terminology.
Ultimately, the ultimate role of AIs in troubleshooting is to connect a novice's inquiry with relevant social annotation accumulated in online resources. 
In that sense, it is crucial to further develop how the AI can work better to elaborate on a user's situation and connect to the relevant solutions.
A mode for representing one's representation to the system must be carefully designed based on the deep consideration of the nature of the target task. We introduce two new directions. 
\begin{itemize}
    \item \textbf{Vision-based Human-AI Loop}:
    Rapid progress in computer vision has enabled vision-based AI models to outperform humans in many tasks. 
    However, when it comes to the way we interact with vision-based models, the way we interact is one-time use rather than iterative. 
    We are at the early stage of designing a feedback loop between humans and vision-based models~\cite{gao2022aligning, gao2022going, sun2023designing}.
    We expect that devising a new form of communication loop that can realize the elaboration and adjustment of a vision-based question can result in opening an intriguing design space in troubleshooting. 
    As of early work, Attention Branch Network~\cite{fukui2019attention} and Convolutional Human-in-the-Loop~\cite{mitsuhara2019embedding} have been introduced in the field of computer vision. A recent study in CSCW started to apply such a feedback loop in HCI and CSCW domains~\cite{gao2022aligning}.
    \item \textbf{Multimodal Interaction between Human and AI}:
    Recent progress in Visual Question Answering (VQA) has developed powerful model architectures that combine images and texts~\cite{zellers2019recognition}.
    While we expect such multi-modal-based communication designs can open a new way to interact with systems for a troubleshooting task and beyond, we identified relatively little research related to this direction in HCI.
\end{itemize}

\subsection{Social Annotation for Scalable 3DPFIX Dataset} \label{section:pipeline}

We initially built our \system dataset for classifying the cases based on the 27 base documents defined by Simplify3D Print Quality Guide~\cite{simplify3d}.
This raises an important question; are the failure types that the Simplify3D guide presented comprehensive enough to capture contemporary trends of printing failures?
As observed in the actual practice, there likely emerges new failure types, terminologies, and visual features as printing technology evolves with new printers and advanced slicing algorithms resolve known major printing issues (e.g., Ultimaker Arachne).
Therefore, renowned online archives may fail to provide inclusive support cases timely. 
To verify this assumption, we first examined the recent subset of community discourses by retrieving \textit{1,000 recent images} from the FixMyPrint dataset, and manually annotated the images by checking visual cues and corresponding comments. The annotation was done by the first author who has extensive knowledge of 3D printing failures and verified by another author who is a 3D printing expert with over 10 years of experience. 
We discovered the 20 most recent common failure types reported by community members:
\begin{itemize}
\item[] \textcolor{blue}{under-extrusion, stringing, blobs, [z-seam], layer shifting, warping}, 
\item[] overheating, vibration, bed leveling, too close/far print nozzle, over-extrusion, layer separation \& [delamination], gaps between infill and outer wall, visible lines on the top, bridging, weak infill, 
\item[] \textcolor{red}{[spaghetti, z-binding, pillowing, elephant foot]}
\end{itemize}
Failure types covered by the initial 3DPFIX dataset are highlighted in \textcolor{blue}{blue}. The initial dataset covers 6 types (having z-seam as a subclass of blobs) of types above, about 30\% of failure types recently reported by the community. 
Aligning with our assumption, we found 4 new cases and 6 new terminologies (highlighted in \textcolor{red}{red} and [brackets], respectively) which are not a part of the Simplify3D guide. 
This gap between static guides and actual cases with the growing need of remote novices motivated us to build a pipeline, scalable and generalizable, that extracts domain knowledge accumulated in online communities from social annotations and constructs a large dataset of images for automated detection and solution pool accumulated by community users. 
More importantly, the pipeline approach eventually reduces manual human labor needed for labeling data, making our system sustainable with new cases powered by human-AI collaboration in the future.
\system dataset pipeline (see Figure~\ref{figure:pipeline} in Appendix~\ref{appendix} for a high-level view) consists of three steps:
\begin{enumerate}[leftmargin=.3in]
\item Visual feature dictionary: from the subset of online posts database, defining a dictionary of failure types and visual features using comments (i.e., social annotation)
\item Failure image dataset: expanding the dataset by annotating posts by referencing the visual feature dictionary \& comments
\item Solution pool: extracting solutions from annotated posts
\end{enumerate}
In addition to what is already shown in Figure~\ref{figure:visual_feature}, we present a failure type dictionary with distinct visual features including all 20 types of printing failure spotted in Figure~\ref{figure:dictionary} in Appendix \ref{appendix}. 
We envision that this can be used as a useful reference to expand the 3DPFIX dataset for new emerging 3D printing issues, for example, through human annotations on a minimal set of images along with state-of-art ML techniques, such as few-shot learning approaches (e.g., \cite{hu2022pushing}) that effectively trains the model with a large unannotated dataset and a few annotated examples. 

\subsection{Longitudinal Deployment Study}
In the S2 setting, we provided a specific error type to participants and asked them to solve the problem. 
To compensate for artificiality in the lab settings, we deployed the system for a short period of time and received comments from real users with their own failures. This enabled us to obtain feedback under less-artificial settings. 
As a future work, we plan to conduct a deployment study to improve remote novices' troubleshooting experience in the long run, to better understand \system's ability to provide reliable \& scalable failure diagnosis and solution suggestions powered by \system pipeline, in realistic situations.

\subsection{Alternative Design Choices \& Multimodal dataset}
In this work, we were able to collect invaluable comments and suggestions to improve the design of \system through S2. One user from S2's mini-deployment mentioned that having more visual aids in presenting solutions would help users better understand the concepts.
Similarly, in future work, we will actively collect feedback from real users and keep improving the design of \system.
Also, we did not disclose the numerical representation (e.g., 90\% probability) but only showed categorical information (e.g., highly likely) in providing AI's decision rationale, according to previous work's observation~\cite{xie2020chexplain} that numerical values can confuse novices. Considering that AIs become more prevalent and also numerical values can be beneficial for educational purposes, we will consider having the option to switch to numerical representations for our future deployment study.
Currently, 3DPFIX considers visual information (image) and text (solution). For example, instead of having sound data in the dataset, the current design encourages users to consider a `clue'. For example, if they heard a popping noise during printing, the filament likely needs to be dry. Not only limited to visual and textual information, we believe that building a multi-modal dataset, such as audio data along with visual/text. Collecting videos instead of images would be an effective way to collect sound data as well, which can provide additional evidence in diagnosing failures.

\section{Conclusion}

In this paper, we designed \system to support remote novices' troubleshooting experience based on the design requirements identified through S1. 
In implementing \system, we leveraged social annotation accumulated in online resources, hoping that such an approach can improve the way remove novices resolve issues more efficiently and effectively.
Our S2 showed that participants' task efficiency, task effectiveness, and learnability-related performance are significantly higher when using the \system than when relying on their current practice.
We hope this work can motivate engendering a full-fledged system that can improve people's realistic troubleshooting practice in the 3D printing domain and beyond.

\section*{Acknowledgement}
This work is partially funded by National Science Foundation, IIS-2213842.



\bibliographystyle{ACM-Reference-Format}
\bibliography{main}

\newpage

\appendix

\section{Appendix} \label{appendix}


\begin{figure*}[h]
    \centering
    \includegraphics[width=0.83\linewidth]{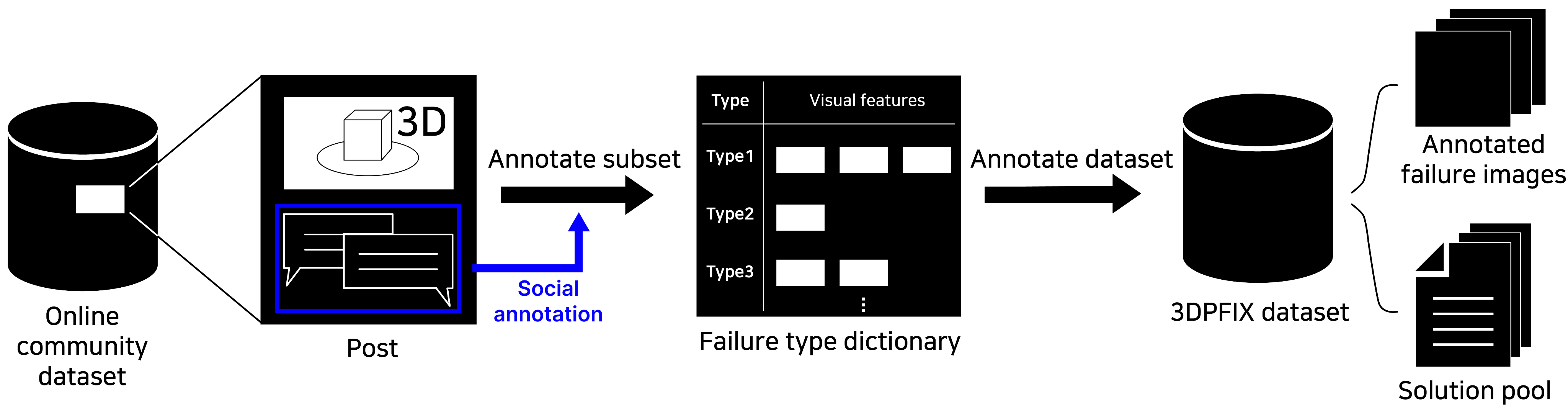}
    \caption{\IssueOneA{3DPFIX dataset building pipeline that leverages social annotation (comments).}}
    \Description{}
    \label{figure:pipeline}
\end{figure*}


\begin{figure*}[h]
    \centering
    \includegraphics[width=0.75\linewidth]{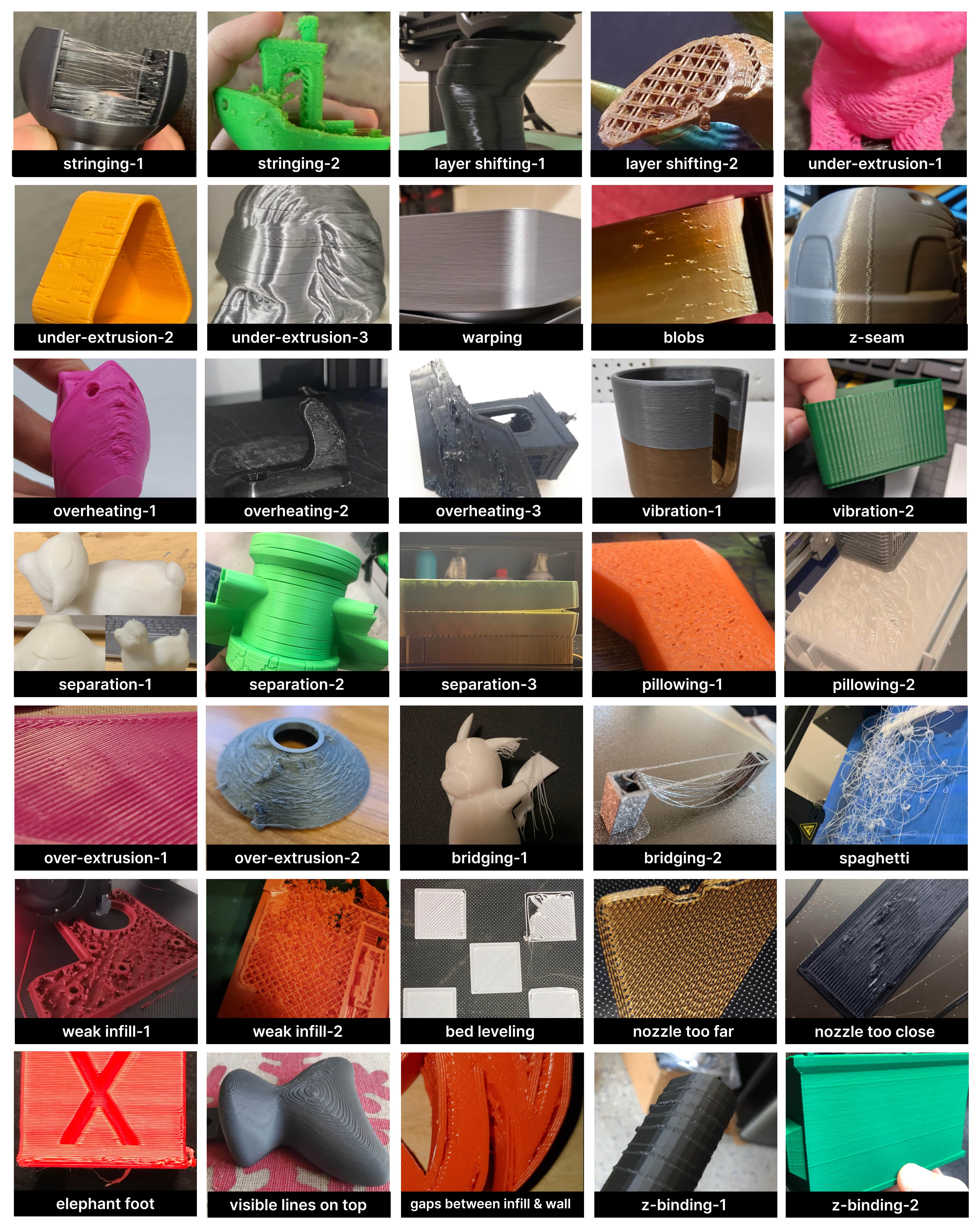}
    \caption{\IssueOneA{Failure type dictionary for 20 failure types spotted from recent posts in the online community.}}
    \Description{}
    \label{figure:dictionary}
\end{figure*}

\received{January 2023}
\received[revised]{July 2023}
\received[accepted]{November 2023}

\end{document}